\documentclass[hidelinks,journal,comsoc]{IEEEtran}
\usepackage{cite}
\usepackage{bm}
\usepackage{amsmath}
\usepackage{extarrows}
\usepackage{amssymb}
\usepackage{graphicx}
\usepackage{color}
\usepackage{enumerate}
\usepackage{bookmark} 
\graphicspath{{figure}}
\usepackage{setspace}
\usepackage{algorithm}
\usepackage{algorithmic}

\newcommand{\mb}{\bm}
\newcommand{\mr}{\mathrm} 
\newcommand{\ms}{\mathrm}
\newcommand{\BE}{\begin{equation}}
\newcommand{\EE}{\end{equation}}
\newcommand{\BS}{\begin{subequations}}
\newcommand{\ES}{\end{subequations}}
\renewcommand{\bf}{\bm}

\newtheorem{theorem}{Theorem}
\newtheorem{proposition}{Proposition}
\newtheorem{assumption}{Assumption}

\newtheorem{lemma}{Lemma}

\ifCLASSINFOpdf
\graphicspath{{figure/}}
   
\else
  
\fi

\begin{document}
 
\title{Capacity Optimality of AMP in Coded Systems}
 
\author{\IEEEauthorblockN{Lei~Liu, \emph{Member, IEEE}, Chulong~Liang, Junjie~Ma, and~Li~Ping, \emph{Fellow, IEEE}}

\thanks{Lei~Liu was with the Department of Electronic Engineering, City University of Hong Kong (CityU), Hong Kong, SAR, China, and is currently with the School of Information Science, Japan Institute of Science and Technology (JAIST), Nomi 923-1292, Japan (e-mail: leiliu@jaist.ac.jp).}
\thanks{Chulong~Liang was with the Department of Electrical Engineering, City University of Hong Kong, Hong Kong SAR, China. He is now with the Algorithm Department, ZTE Corporation, Shenzhen 518057, China, and also with the State Key Laboratory of Mobile Network and Mobile Multimedia Technology, ZTE Corporation, Shenzhen 518057, China (e-mail: liang.chulong@zte.com.cn).\par
Junjie Ma is with the Institute of Computational Mathematics and Scientific/Engineering Computing, Academy of Mathematics and Systems Science, Chinese Academy of Sciences, and also with University of Chinese Academy of Science, Beijing, China (e-mail: majunjie@lsec.cc.ac.cn).\par
Li~Ping is with the Department of Electronic Engineering, City University of Hong Kong, Hong Kong, SAR, China (e-mail: eeliping@cityu.edu.hk).\par
}
}

\maketitle

\begin{abstract}
This paper studies a large random matrix system (LRMS) model involving an arbitrary signal distribution and forward error control (FEC) coding. We establish an area property based on the approximate message passing (AMP) algorithm. Under the assumption that the state evolution for AMP is correct for the coded system, the achievable rate of AMP is analyzed. We prove that AMP achieves the constrained capacity of the LRMS with an arbitrary signal distribution provided that a matching condition is satisfied. As a byproduct, we provide an alternative derivation for the constraint capacity of an LRMS using a proved property of AMP. We discuss realization techniques for the matching principle of binary signaling using irregular low-density parity-check (LDPC) codes and provide related numerical results. We show that the optimized codes demonstrate significantly better performance over un-matched ones under AMP. For quadrature phase shift keying (QPSK) modulation, bit error rate (BER) performance within 1 dB from the constrained capacity limit is observed. 
\end{abstract}


\IEEEpeerreviewmaketitle

\section{Introduction}
\subsection{Large Random Matrix System (LRMS) and Approximate Message Passing (AMP)}
Consider the problem of signal reconstruction for a large random matrix system (LRMS):
\BE\label{Eqn:linear_system}
\bf{y}=\bf{Ax}+\bf{n} 
\EE
where $\bf{A}$ is an $M\times N$ matrix with independent and identically distributed (IID) entries and $\bf{x}$ a length-$N$ vector with IID entries. Furthermore, we assume that the entries of $\bf{A}$ are Gaussian, but those of $\bf{x}$ are not necessarily Gaussian. If $\bf{x}$ is generated using a forward error control (FEC) code with rate $R_{\cal C}$, the overall rate of this scheme (per received symbol) is $NR_{\cal C}/M$.

In a special case when $\bf{x}$  is un-coded, if $\bf{x}$ is Gaussian, the optimal solution can be obtained using the linear minimum mean square error (MMSE) methods. Otherwise, the problem is in general NP hard \cite{Micciancio2001,verdu1984_1}. Approximate message passing (AMP), derived from belief-propagation (BP) with Gaussian approximation and first order Taylor approximation, has attracted extensive research interest for this problem \cite{Bayati2011, Donoho2009}. A basic assumption of AMP is that $\bf{A}$ has IID Gaussian (IIDG) entries. This assumption will hold throughout this paper.

 AMP works by iterating between two local processors: namely, a linear detector (LD) and a non-linear detector (NLD). (An NLD is sometimes referred to as a de-noisier \cite{Schniter2016learning}.) There is no matrix inversion involved, so its complexity is low. AMP has been studied for various signal processing \cite{ Schniter2016learning, Kamilov2014, Som2012, Schniter2012phase, Ma2014denoising, kabashima2003cdma, neirotti2005improved, Wu2014AMP, Nassar2014OFDM, Schniter2011,Nabaee2014quantized, Rush2017SSC, Liang2020, Barbier2017SSC} and communication applications \cite{Liang2017CL, Liang2016ISTC}. Recently, it has been observed that AMP and its variations such as expectation propagation (EP) \cite{Cakmak2018, Minka2001} and orthogonal AMP (OAMP) \cite{Ma2016} outperform the conventional Turbo linear MMSE (Turbo-LMMSE) in coded linear systems involving FEC coding \cite{Santos2017, MengVTC2015, MaTWC}. The applications of such systems include inter-symbol interference (ISI) channels \cite{Santos2017}, multi-user systems \cite{MengVTC2015}  and multiple-input multiple-output (MIMO) systems \cite{MaTWC}. Most works on AMP in coded systems are simulation based \cite{Santos2017, MengVTC2015, MaTWC}. There is still a lack of rigorous analysis on the information theoretical limits of AMP in coded systems.

\subsection{Contributions of this Paper}
In this paper, we discuss the LRMS in \eqref{Eqn:linear_system} with FEC coding. The receiver is a variation of AMP with NLD formed by an \emph{a posteriori probability} (APP) decoder. For convenience of discussions, we define two classes of optimality for a receiver. 
\begin{itemize}
  \item A receiver is MMSE-optimal if it can achieve MMSE when $\bf{x}$ is an IID sequence.
  \item A receiver is information theoretically optimal if it can achieve error free performance when $\bf{x}$ is coded with a rate which equals to the mutual information $I(\bf{x}; \bf{y})$. 
\end{itemize}

The state evolution (SE) technique of AMP was originally derived to track the mean square error (MSE) in AMP during iterative processing. SE involves a scalar recursion of the transfer functions of LD and NLD. It has been shown via SE analysis that AMP can achieve MMSE asymptotically in the un-coded case when the transfer functions of LD and NLD have only one fixed-point \cite{ Tulino2013,  Barbier2017arxiv, Reeves_TIT2019}. In this paper, we will show via SE analysis that AMP is information theoretically optimal, while the conventional methods, such as the well-known Turbo-MMSE algorithm  \cite{Lei20161b,Yuan2014} (also referred to as the Wang and Poor algorithm \cite{Wang1999,MaTWC,Loeliger2007}), are not.  

Our discussions are based on the following background works: (i) the I-MMSE relationship between mutual information and MMSE  \cite{Guo2005}, (ii) the area property of iterative decoding systems \cite{Bhattad2007}, (iii) the MMSE-optimality of AMP \cite{Tulino2013,  Barbier2017arxiv, Reeves_TIT2019}, and (iv) the capacity of an LRMS recently derived in \cite{Barbier2017arxiv, Barbier2016conf, Reeves_TIT2019}. Similarly to \cite{Yuan2014, Lei20161b }, the performance of AMP can be optimized by matching the transfer functions of LD and decoder. The achievable rate can be analyzed using an area property similar to that for low density parity check (LDPC) decoders \cite{Yuan2014, Lei20161b}. However, there is a key difference. The area property for LDPC decoders is based on the so-called extrinsic information, for which perfect matching is theoretically possible \cite{Yuan2014, Lei20161b}. We will see that perfect matching is not possible for AMP: there is an inherent gap between the two transfer functions. Interestingly, AMP is still information theoretically optimal despite this gap, in the sense that its achievable rate can approach the mutual information $I(\bf{x}; \bf{y})$.

The main contributions of this paper are summarized as follows.
\begin{itemize}
  \item We show that the constrained capacity of a coded LRMS with an arbitrary input distribution (Gaussian or non-Gaussian) can be graphically interpreted as the area determined by the transfer functions of LD and MMSE NLD of an AMP. We establish an area property for AMP and derive its achievable rate under a matching condition. We prove that this achievable rate equals to the constrained capacity of an LRMS derived in \cite{Barbier2017arxiv, Barbier2016conf, Reeves_TIT2019}, thereby showing the potential information theoretic optimality of AMP in coded linear systems.  
  
  \item  We develop a matching strategy for AMP. We show the existence of a capacity approaching superposition coded modulation (SCM) scheme for Gaussian signaling. We also provide numerical results to demonstrate the efficiency of the matching strategy for binary signaling. These findings provide a promising direction to significantly enhance the performance of coded linear systems.  
  
  \item As a byproduct, we provide an alternative derivation for the capacity of an LRMS. This capacity has been recently derived in \cite{Barbier2017arxiv, Barbier2016conf, Reeves_TIT2019}. In our opinion, the approach in this paper is more concise, taking advantage of the available results of AMP. 
\end{itemize}

\subsection{Connection to Existing Works}
In \cite{Reeves_TIT2019}, the authors derived the constrained capacity and MMSE of an LRMS by establishing some properties of the finite-length MMSE and mutual information sequences, and then using these properties to uniquely characterize their limits.
The authors of \cite{Barbier2017arxiv, Barbier2016conf} provided a rigorous proof for the replica formula of the constrained capacity (see Theorem \ref{Pro:dis_cap}) by using a Guerra-Toninelli type interpolation method to yield an upper bound for the capacity, and spatial-coupling and AMP to yield a lower bound.  In addition, the MMSE optimality of AMP was rigorously proved in \cite{Barbier2017arxiv, Barbier2016conf}. In this paper, we give a different concise derivation of the capacity. Our derivations are built on existing results, namely, the I-MMSE theorem \cite{Guo2005, Bhattad2007}, the MMSE optimality of AMP \cite{Tulino2013,  Barbier2017arxiv, Reeves_TIT2019} and the decoupling property of AMP \cite{Takeuchi2019AMP}. 

Sparse regression coding (SRC) \cite{Rush2017SSC, Barbier2017SSC} is a special case of \eqref{Eqn:linear_system} in which $\bf{x}$ is generated using position modulation. A position modulation scheme of block length $B$ is a special form of FEC coding with coding rate $R_{\rm PM} = \log_2(B)/B$, which is also equivalent to a length-$B$ Hadamard code \cite{Liang2016ISTC, Liang2020}. The decoding technique for SRC \cite{Rush2017SSC, Barbier2017SSC} can be regarded as a special form of AMP, with FEC decoding implemented by position demodulation. SRC is capacity approaching when $B\to \infty$ or, equivalently, $R_{\rm PM}\to0$. (Note: The overall rate $R_{\rm PM}\times N/M$ of SRC can remain finite even though $R_{\rm PM}\to0$ if $N/M\to \infty$.) The cost of position demodulation grows with $B$. For affordable complexity (e.g. $B$ up to thousands), there is a considerable gap between SRC performance and capacity. Detailed discussions on this issue can be found in \cite{Liang2020}. 

The compressed coding scheme introduced in \cite{Liang2020} is equivalent to the LRMS in \eqref{Eqn:linear_system}. Arbitrary FEC coding is assumed in \cite{Liang2020}. The achievable rate of compressed coding using AMP approaches Gaussian capacity in the limiting case when the underlying FEC rate $R_{\cal C}\to0$ \cite{Liang2016ISTC, Liang2020}, which is similar to SRC \cite{Rush2017SSC, Barbier2017SSC}. It is shown that compressed coding using a properly designed FEC code can outperform SRC under practical complexity constraint (i.e., a limited block length of position modulation in SRC) \cite{Liang2020}. 

Recall that the overall rate of the LRMS in \eqref{Eqn:linear_system} is $NR_{\cal C}/M$. It was proved in \cite{Rush2017SSC, Barbier2017SSC, Liang2020} that SRC and compressed coding are capacity approaching when $R_{\rm PM}\!\to\!0$ or $R_{\cal C}\!\to\!0$. In these cases, if the overall rate is finite, we need $N/M\!\to\!\infty$ in \eqref{Eqn:linear_system}, which may incur excessively high receiver cost. In this paper, we remove this limitation. We will show that AMP is information theoretically optimal for any $R_{\cal C}$ under a matching condition.

\subsection{Notations}
Boldface lowercase letters represent vectors and boldface uppercase symbols denote matrices. $I({\bf{\bf{x}};{\bf{y}}})$ for the mutual information between $\bf{x}$ and $\bf{y}$, $\mb{I}$ for the identity matrix with a proper size, $|\cal{S}|$ for the cardinality of set $\cal{S}$, $\bm{a}^{\mr{H}}$ for the conjugate transpose of $\bm{a}$, $\|\bm{a}\|$ for the $\ell_2$-norm of the vector $\bm{a}$, $\det(\bf{A})$ for the determinant of $\bf{A}$, $\ms{Tr}(\bm{A})$ for the trace of $\bm{A}$, $A_{ij}$ for the $i$th-row and $j$th-column element of $\bf{A}$, $\mathcal{CN}(\bm{\mu},\bm{\Sigma})$ for the circularly-symmetric Gaussian distribution with mean $\bm{\mu}$ and covariance $\bm{\Sigma}$, $\ms{E}\{\cdot\}$ for the expectation operation over all random variables involved in the brackets, except when otherwise specified. $\mr{E}\{a|b\}$ for the expectation of $a$ conditional on $b$, $\ms{var}\{{a}\}$ for $\ms{E}\big\{ \!\left({a} - \ms{E}\{{a}\}\right)^2 \big\}$, $\ms{mmse}\{{a}|{b}\}$ for $\ms{E}\left\{ \left({a} \!-\! \ms{E}\{{a}|{b}\}\right)^2|{b} \right\}$, $\langle\bf{x}\rangle \!=\! \sum_{i=1}^N x_i/N$, and $\eta'(r)\!=\!\frac{\partial}{\partial r} \eta(r)$.    

{Capacity is defined by default as the maximum mutual information over all possible choices of input distribution. A constrained capacity is defined as the mutual information under a fixed input distribution $x\sim P_X(x)$. For the systems considered in this paper, capacity is always achieved by Gaussian signaling. Hence sometimes we will call it ``Gaussian capacity". For simplicity, we will call the constrained capacity given $P_X(x)$ ``capacity" if it is clear by context.}
\subsection{Paper Outline}
This paper is organized as follows. Section II gives the area property, the constrained capacity of an LRMS and the  AMP algorithm. Section III proves the capacity optimality of AMP under Gaussian assumption. Numerical results are shown in Section IV.

\section{Preliminaries} 
\begin{figure}[t]
  \centering
  \includegraphics[width=6cm]{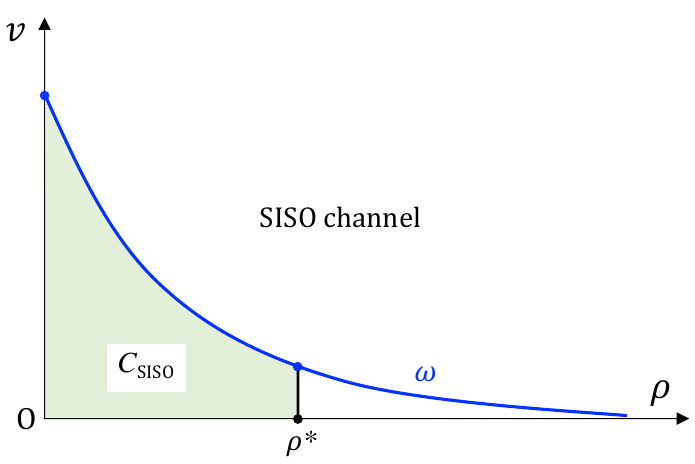}\\
  \caption{Graphical illustration of the capacity of a SISO-AWGN channel. $\rho^*$ denotes the channel SNR, and $\omega$ is the MMSE of the SISO-AWGN channel. }\label{Fig:AWGN_area}
\end{figure}

{In this section, we briefly outline some existing results that will be used in this paper, such as the area property for single-input-single-output additive white Gaussian noise (SISO-AWGN) channel, the capacity of a large random matrix system, and the AMP algorithm.

\subsection{Area Property for SISO-AWGN Channel}\label{Sec:SISO_area}
A SISO-AWGN channel is defined as \BE\label{Eqn:SISO_AWGN}
y=\sqrt{\rho}x+z,
\EE
where $x\sim P_X(x)$, $z\sim \mathcal{CN}(0,1)$, and $\rho$ denotes the signal-to-noise-ratio (SNR). The MMSE of \eqref{Eqn:SISO_AWGN} is denoted as
\BE
\omega(\rho) \equiv \mr{mmse}(x|\sqrt{\rho}x+z, x\sim P_X(x)).
\EE

The following lemma, proved in \cite{Guo2005}, establishes the connection between MMSE and the capacity given $P_X(x)$ for a SISO-AWGN channel.

\begin{lemma}[Scalar I-MMSE]\label{Lem:S-I-MMSE}
Let SNR$=\rho^*$. The capacity of a SISO-AWGN channel equals to the area under  $\omega(\rho)$ from $\rho=0$ to $\rho=\rho^*$, i.e., 
\BE\label{Eqn:C_mmse}
C_{\rm SISO}(\rho^*) = I\big({x}; \sqrt{\rho^*}x+z\big) = \int_{0}^{\rho^*} \omega(\rho) d\rho.
\EE
\end{lemma}

Fig. \ref{Fig:AWGN_area} gives a graphical illustration of Lemma \ref{Lem:S-I-MMSE}. The following are some instances of the MMSE function $\omega(\rho)$. 
\begin{itemize}
    \item  \emph{Gaussian Signaling:} For $x\sim\mathcal{CN}(0,1)$, $\omega(\rho)$ is given by
\BS\BE
 \omega_{\mr{Gau}}(\rho)\equiv{1}/({\rho+1}).
\EE
The channel capacity with Gaussian signaling is given by
\BE 
 C^{\rm Gau}_{\rm SISO} \!= \!\! \! \int_{0}^{\rho^*} \!\!\!\!\!\omega_{\mr{Gau}}(\rho) d\rho =\!\! \int_{0}^{\rho^*} \!\!\!\! \!\frac{1}{1\!+\!\rho} \,\, d\rho = \log(1\!+\!\rho^*).
\EE\ES
\item \emph{Discrete Signaling \cite{Lozano2006}:} For an arbitrary discrete constellation $\mathcal{S}=\{s_1,\cdots,s_{|\mathcal{S}|}\}$ with equal probability ${1}/{|\mathcal{S}|}$, $\omega(\rho)$ is given by
\BS\BE\label{Eqn:QAM_MMSE}
 \omega_{\mathcal{S}}(\rho)\equiv1-\frac{1}{\pi} \int\frac{\left|\sum_{l=1}^{|\mathcal{S}|}s_le^{-|y-\sqrt{\rho}s_l|^2}\right|^2} {|\mathcal{S}| \sum_{l=1}^{|\mathcal{S}|}e^{-|y-\sqrt{\rho}s_l|^2} } d y.
\EE
The constrained capacity for $x\in \mathcal{S}$ is given by $\int_{0}^{\rho^*} \!\!\!\omega_{\mathcal{S}}(\rho) d\rho$. In particular, if $\bf{x}$ is un-coded, the information rate is 
\BE
R_{\mathcal{S}}= \int_{0}^{\infty} \!\!\!\omega_{\mathcal{S}}(\rho) d\rho =\log |\mathcal{S}|.
\EE\ES

\item \emph{QPSK Signaling \cite{Guo2005}:} As a special case of \eqref{Eqn:QAM_MMSE}, for quadrature phase-shift keying (QPSK) signaling $x\in\{\frac{1}{\sqrt{2}}(\pm 1\pm j)\}$, $\omega(\rho)$ is given by
\BE\label{Eqn:QPSK_MMSE}
 \omega_{\mr{QPSK}}(\rho)\equiv1-\int_{-\infty}^{\infty}\frac{e^{-y^2/2}}{\sqrt{2\pi}} \mr{tanh}(\rho-\sqrt{\rho}y) dy.
\EE
\item \emph{Code-Rate-MMSE Lemma \cite{Bhattad2007}:} Let the code length be $N$ and code rate $R=K/N$. We treat the code-book $\bf{{\mathcal{C}}}=\{\bf{c}_1,\cdots,\bf{c}_{2^K}\}$ as a uniformly distributed $N$-dimension constellation with ${2^K}$ discrete points. When SNR$\to \infty$, the capacity per length-$N$ code block approaches to the entropy of $\bf{{\mathcal{C}}}$, i.e., $\log(2^K)=K$. The entropy per dimension is $K/N$. Hence, we have
\BE\label{Eqn:R_mmse}
R_{\cal C} =  \int_{0}^{\infty}\omega_{\mathcal{C}}(\rho)d\rho =K/N,
\EE
where  $ \omega_\mathcal{C}(\rho) \equiv \tfrac{1}{N}\mr{mmse}(\bf{x}|\sqrt{\rho}\bf{x}+\bf{z},{\bf{x}\in {\bf{{\mathcal{C}}}}})$ and $R_{\cal C} $ is the rate of $\bf{{\mathcal{C}}}$. 
\end{itemize} }

\subsection{LRMS Capacity}
\begin{figure}[t] 
  \centering
  \includegraphics[width=7cm]{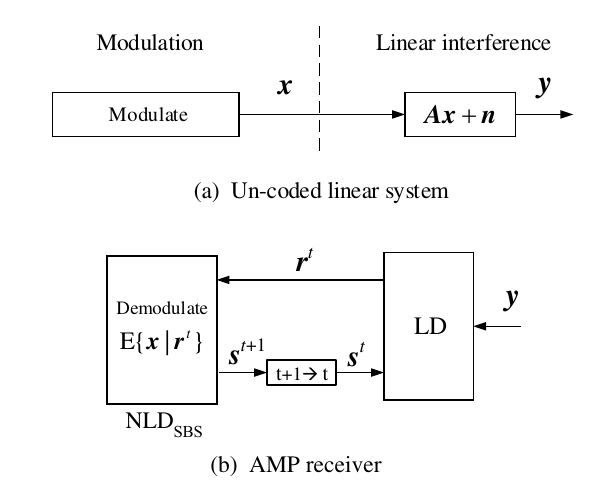}\\  
  \caption{Un-coded LRMS: transmitter and AMP receiver, where ``Demodulate'' and LD in (b) correspond to ``Modulate'' and ``$\bf{A}\bf{x}+\bf{n}$'' in (a) respectively.}\label{Fig:model} 
\end{figure}

Return to the LRMS in \eqref{Eqn:linear_system}: $\bf{y}=\bf{Ax}+\bf{n}$,
where $\bm{y}\!\in\!\mathbb{C}^{M\times1}$ is a vector of observations, $\bf{A}\!\in\!\mathbb{C}^{M\times N}$ an IIDG matrix with $A_{ij}\sim \mathcal{CN}({0},1/M)$\footnote{In fact, it can be easily extended to a more general case $A_{ij}\sim \mathcal{CN}({0},\sigma^2_a/M)$, where $\sigma^2_a$ is finite. In this case, we can rewrite the system to ${\bf{y}}' =\sigma_a^{-1}{\bf{y}} = {\bf{A'x}} +{\bf{n}}'= \sigma_a^{-1}{\bf{Ax}} + \sigma_a^{-1}{\bf{n}}$, where $ A'_{ij}\sim \mathcal{CN}({0},1/M)$ and $\bm{n}'\!\sim\!\mathcal{CN}(\mathbf{0},\sigma^2\sigma_a^{-2}\bm{I})$. Then, all the results in this paper are still valid by replacing $\sigma^2$ with $\sigma^2\sigma_a^{-2}$. For example, if $A_{ij}\sim \mathcal{CN}({0},1/N)$, we replace $\sigma^2$ by $\beta\sigma^2$ to make the results of this paper be valid.}, $\{x_i\sim P_X(x), \forall i\}$, and $\bm{n}\!\sim\!\mathcal{CN}(\mathbf{0},\sigma^2\bm{I}_M)$ a vector of Gaussian additive noise samples. Fig. \ref{Fig:model}(a) shows 
a modulated LRMS. In this paper, we consider a large system with $M,N\to\infty$ and a fixed $\beta=N/M$. The transmit SNR is defined as $snr = {\mr{E}\{\|x_i\|^2\}}/{\mr{E}\{\|n_j\|^2\}} = \sigma^{-2}$.
We assume that $\bf{A}$ is known at the receiver, but unknown at the transmitter\footnote{If $\bf{A}$ is also known at the transmitter, then the LRMS in \eqref{Eqn:linear_system} can be converted to a set of parallel SISO-AWGN channels using singular value decomposition (SVD). Then, the well-known water filling technique is capacity approaching.}. This assumption has been widely used in multiple-input-multiple-output (MIMO) and multi-user MIMO (MU-MIMO) systems \cite{ Lei20161b, MaTWC, MengVTC2015}.

The constrained capacity of an LRMS given $P_X(x)$ was proved in \cite{Barbier2017arxiv, Barbier2016conf, Reeves_TIT2019}.

\begin{lemma}[Capacity]\label{Pro:dis_cap}   
Assume that the signal distribution $P_X(x)$ satisfies the single-crossing property, i.e., $\zeta =\beta \, snr \, \omega\big(1/[\beta(1+\zeta)]\big)$ has exactly one positive fixed point $\zeta^*$. Then, the capacity of the LRMS in \eqref{Eqn:linear_system} is given by \vspace{-0.1cm}
\BE\label{Eqn:dis_cap}
C  \!=\!  \beta^{-1}\big[ \log({1\!+\!\zeta^*})-{\zeta^*}\!/({1\!+\!\zeta^*}) \big] + C_{\rm SISO} \left({snr}/(1\!+\!\zeta^*)\right),  
\EE 
where $C_{\rm SISO}(\cdot)$ is defined in \eqref{Eqn:C_mmse}.
\end{lemma}

The capacity in \eqref{Eqn:dis_cap} is equivalent to that desired in \cite{Reeves_TIT2019}. For the details, see APPENDIX \ref{APP:Consistency}.

\subsection{Overview of AMP}\label{Sec:AMP_uncoded}
AMP \cite{Donoho2009}  finds an approximate MMSE solution to the problem in \eqref{Eqn:linear_system} using the following iterative process (initialized with $t=0$ and $\bf{s}^0=\bf{r}^0_{\mr{Onsager}}=\bf{0}$): 
\BS\label{Eqn:AMP}\begin{align}
 \mathrm{LD:}\;\; & \bf{r}^t\!=\! f(\bf{s}^t) \!\equiv \! \bf{s}^t \!+\! \bf{A}^{\rm H}(\bf{y}\!-\!\bf{A}\bf{s}^t) \!+\! \bf{r}^t_{\mr{Onsager}},\label{Eqn:LD}\\
\mathrm{NLD_{SBS}}: \;\; &\bf{s}^{t+1} = \eta(\bf{r}^{t})\equiv \mr{E}\{\bf{x}|\bf{r}^{t}\},\label{Eqn:NLD}
\end{align}\ES
where $\eta(\bf{r}^{t})$ is a symbol-by-symbol (SBS) MMSE demodulate function, and $\bf{r}^t_{\mr{Onsager}}$ is an ``Onsager term''  defined by $\bf{r}^t_{\mr{Onsager}}\!=\!\beta \langle\eta'(\bf{r}^{t-1})\rangle (\bf{r}^{t-1}\!-\!\bf{s}^{t-1})$ \cite{Donoho2009}. Fig. \ref{Fig:model}(b) is a graphical illustration of AMP, where the linear detector (LD) and non-linear detector (NLD) correspond to \eqref{Eqn:LD} and \eqref{Eqn:NLD} respectively. 
We define the errors at the LD and NLD respectively as
\BE\label{Eqn:errors}
\bf{h}^t \equiv \bf{r}^t -\bf{x} \quad {\rm and} \quad  \bf{q}^t \equiv \bf{s}^t -\bf{x}.
\EE 
Let $\rho^t$ be the signal-to-interference-plus-noise-ratio (SINR) for $\bf{r}^t$ and $v^t$ the MSE for $\bf{s}^t$:
\BE\label{Eqn:rho_v}
  \rho^t \equiv N\big[{\rm E}\big\{\|\bf{h}^t\|^2\big\}\big]^{-1},\qquad
  v^t \equiv  \tfrac{1}{N}{\rm E}\big\{\|\bf{q}^t\|^2\big\}.
\EE
The following lemma summarizes the findings in \cite{Bayati2011}.  

\begin{lemma}\label{Pro:SE}
Let $M,N\to\infty$ with a fixed $\beta=N/M$. For AMP, $\bf{h}^t$ defined in \eqref{Eqn:errors} can be modeled by a sequence of IIDG samples independent of $\bf{x}$. The LD and NLD of AMP can be characterized by the following transfer functions \cite{Bayati2011}
 \BS \begin{align}
\mathrm{LD:}& \quad \rho^t  =\phi(v^t) = ({\beta v^t + \sigma^2})^{-1},\label{Eqn:LD_form}\\
\mathrm{NLD_{SBS}:}& \quad  v^{t+1}= \omega(\rho^t), \label{Eqn:NLD_MMSE}
\end{align}\ES
where $\omega(\cdot)$ is the MMSE function given in \ref{Sec:SISO_area}. 
\end{lemma}
 
\begin{figure}[t]
  \centering
  \includegraphics[width=8cm]{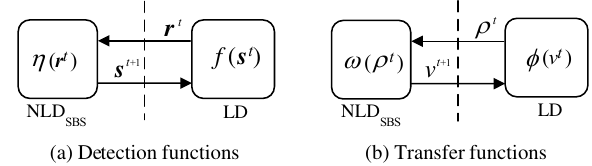}\\ 
  \caption{Detection functions (a) and transfer functions (b).}\label{Fig:SVTF} 
\end{figure}

The iterative process in \eqref{Eqn:AMP} can be written as (see Fig. \ref{Fig:SVTF})
\BS\BE\label{Eqn:ite_pro}
\bf{r}^0  \!=\! f(\bf{s}^0), \; \bf{s}^1  \!=\! \eta(\bf{r}^0),\;\bf{r}^1  \!=\!f(\bf{s}^1), \; \bf{s}^2  \!=\!\eta(\bf{r}^1), \dots
\EE
From \eqref{Eqn:rho_v} and Lemma \ref{Pro:SE}, we can track the SINR and MSE in \eqref{Eqn:ite_pro} as 
\BE
  \rho^0  \!=\!\phi(v^0), \; v^1 \! =\!\omega(\rho^0),\;\rho^1  \!=\!\phi(v^1), \; v^2  \!=\!\omega(\rho^1) \dots
\EE\ES 

 \begin{figure}[t] 
  \centering
  \includegraphics[width=7cm]{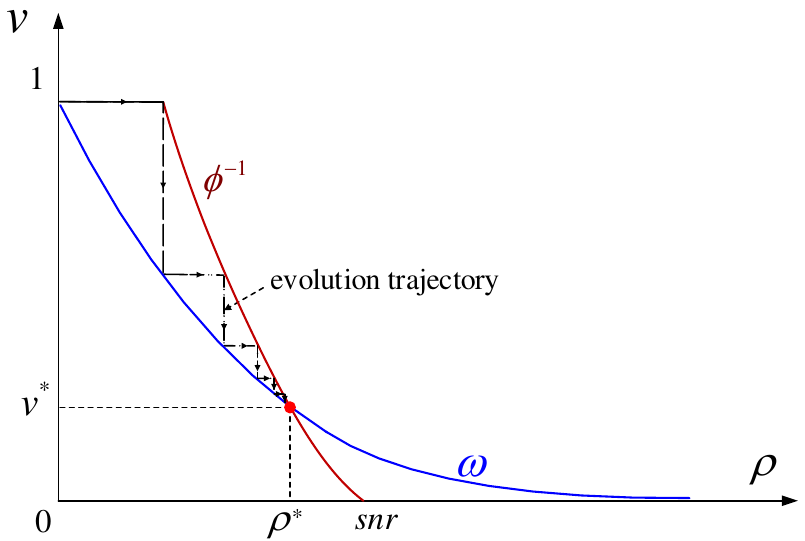}\\
  \caption{Graphical illustration of an  AMP, where $\phi^{-1}$ is the inverse of $\phi$ given in \eqref{Eqn:LD_form} and $\omega$ is defined in \eqref{Eqn:NLD_MMSE}. The iterative process of AMP is illustrated by the evolution trajectory, and the fixed point $(\rho^*, v^*)$ gives the MMSE. From \eqref{Eqn:LD_form}, we have $\phi(0)=snr$. {(Figure parameters: $\omega(\rho)=\omega_{\rm QPSK}(\rho)$, $\beta = 0.65, snr=5\; \mr{dB}, \rho^* = 2.25, v^*=0.20$.)} }\label{Fig:TF_chart} 
\end{figure}

{\begin{assumption}\label{Pro:SCP}
There is exactly one fixed point for $\omega(\rho) = \phi^{-1}(\rho)$ for $\rho>0$, where $\phi^{-1}(\cdot)$ is the inverse of $\phi(\cdot)$.
\end{assumption}}

Fig. \ref{Fig:TF_chart} provides a graphical illustration of Assumption \ref{Pro:SCP}. The evolution trajectory of AMP converges to a unique fixed point $(\rho^*, v^*)$ with $v^*=\omega(\rho^*)$. The following theorem was first established in \cite{Tulino2013} via replica method, and then was rigorously proved in \cite{Barbier2017arxiv, Reeves_TIT2019}.

\begin{lemma}[MMSE Optimality]\label{Lem:mmse}
Let $\hat{{\bf{x}}}_{\mr{MMSE}}=\mr{E}\{\bf{x}|\bf{y},x_i\!\sim\! P_X(x), \forall i\}$ be the conditional mean of $\bf{x}$ given $\bf{y}$ and $\{x_i\!\sim\! P_X(x), \forall i\}$ and suppose  that {Assumption \ref{Pro:SCP} holds.} Then 
\BE\label{Eqn:snr_rho}
v^*=\tfrac{1}{N}\mr{E}\big\{ \|\bf{x}\!-\!\hat{\bf{x}}_{\mr{MMSE}}\|^2 \big\},
\EE
 i.e., AMP converges to the MMSE of the un-coded LRMS. 
\end{lemma}

\section{Capacity Optimality of AMP under Gaussian Assumption}\label{Sec:Cap_Opt} 
In this section, we investigate the achievable rate of the  AMP receiver with FEC decoding.

\subsection{Coded System Model and AMP}
\begin{figure}[t]  
  \centering
  \includegraphics[width=7cm]{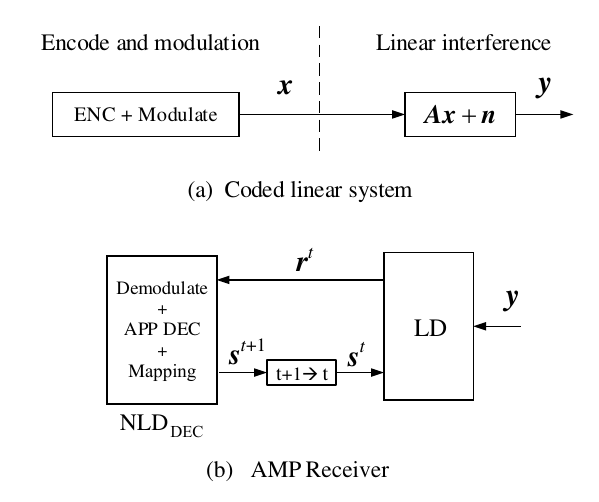}\\
  \caption{Coded linear system: Transmitter and AMP receiver.   ``APP DEC'' (\emph{a-posteriori} probability decoding), ``Demodulate'' and LD in (b) correspond to ``ENC'' (encode),  ``Modulate'' and ``$\bf{A}\bf{x}+\bf{n}$'' in (a) respectively.}\label{Fig:model_coded} 
\end{figure}
Fig. \ref{Fig:model_coded}(a) shows an LRMS involving FEC coding. We write $\bf{x}\in \mathbf{\mathcal{C}}$ for coded $\bf{x}$. The other conditions are the same as that in Fig. \ref{Fig:model}. We focus on the AMP receiver in Fig. \ref{Fig:model_coded}(b) for a coded LRMS.

\emph{\emph{AMP:}} Initialized with $t\!=\!0$ and $\bf{s}^0\!=\bf{r}^0_{\mr{Onsager}}=\bf{0}$,
\BS\label{Eqn:Turbo-AMP}\begin{align}
&\mathrm{LD:}\quad\; \bf{r}^t\!=\! f(\bf{s}^t) \!\equiv \! \bf{s}^t + \bf{A}^{\rm H}(\bf{y}\!-\!\bf{A}\bf{s}^t) + \bf{r}^t_{\mr{Onsager}}, \\
&\mathrm{NLD_{DEC}:} \;\; \bf{s}^{t+1} = \eta_{\cal C}(\bf{r}^{t})\equiv \mr{E}\{\bf{x}|\bf{r}^{t}, {\bf{x}\in {\bf{{\mathcal{C}}}}}\}.\label{Eqn:Turbo-NLD}
\end{align}\ES
Comparing \eqref{Eqn:Turbo-AMP} and \eqref{Eqn:AMP}, we can see that the symbol-wise NLD in AMP {for un-coded $\bf{x}$} is replaced by an \emph{a-posteriori} probability (APP) decoder in AMP {for coded $\bf{x}$}.

Lemma \ref{Pro:SE} gives the IIDG property for AMP for un-coded $\bf{x}$. The discussions in this paper are based on the following assumption for coded $\bf{x}$.  

\begin{assumption}\label{Pro:SE_new}
Lemma \ref{Pro:SE} still holds for AMP for coded $\bf{x}$, i.e., $\bf{h}^t$ is IIDG and independent of $\bf{x}$, and LD and NLD of AMP can be characterized by   
\BS\label{Eqn:Turbo_AMP_SE}\begin{alignat}{2}
  \mathrm{LD:}  &\;\;&&\rho  =\phi(v), \label{Eqn:T_AMP_SE_LD}\\
  \mathrm{NLD_{DEC}:}  &\;\;  &&v= \!\omega_\mathcal{C}(\rho)\!\equiv\! \tfrac{1}{N}\mr{mmse}(\bf{x}|\sqrt{\rho}\bf{x}\!+\!\bf{z},{\bf{x}\!\in\! {\bf{{\mathcal{C}}}}}).\label{Eqn:NLD_MMSE_coded}   
 \end{alignat}
\ES
\end{assumption} 

The $\phi(v)$ in \eqref{Eqn:T_AMP_SE_LD} is the same as that in \eqref{Eqn:LD_form}, and $\omega_\mathcal{C}(\rho)$ depends on the code constraint.  

\subsection{Area Property of LRMS and Capacity Optimality of AMP}\label{Sec:area_AMP} 
In the un-coded case in \eqref{Eqn:AMP}, as shown above, AMP converges to a fixed $(\rho^*,v^*)$ in Fig. \ref{Fig:TF_chart}. Detection is not error free as $v^*>0$. 

\begin{figure}[t] 
  \centering
  \includegraphics[width=7cm]{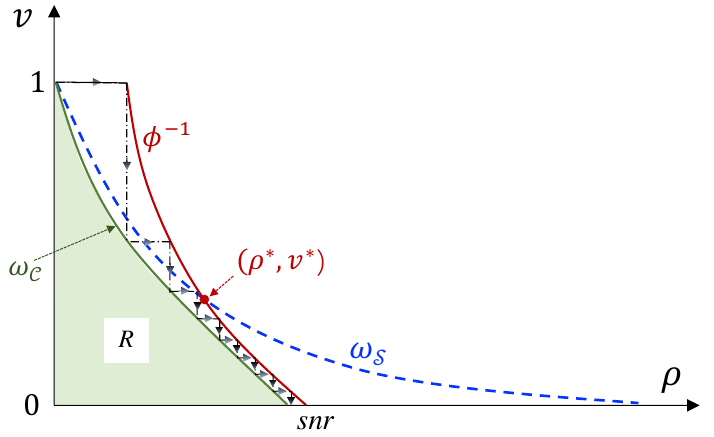}\\ 
  \caption{Graphical illustration of AMP, where $\omega_{\mathcal{S}}$ is a demodulation function (un-coded case) and $\omega_{{\mathcal{C}}}$ is a transfer function of a decoder (coded case). The iterative process of AMP is illustrated by the evolution trajectory between $\phi^{-1}$ and $\omega_{{\mathcal{C}}}$. }\label{Fig:track}
\end{figure}

In the coded case, it is possible to achieve error-free detection using a properly designed $\omega_\mathcal{C}(\rho)$. As illustrated in Fig. \ref{Fig:track}, the key is to create a detection tunnel that converges to $v=0$, implying zero error rate.  There should be no fixed point between $\omega_\mathcal{C}(\rho)$ and $\phi^{-1}({\rho })$, since otherwise the tunnel will be closed at $v>0$. This requires that 
 \BS\BE\label{Eqn:error-free}
\omega_{{\mathcal{C}}}(\rho) <\phi^{-1}(\rho), \;\;\;  {\rm for} \;\;0\le\rho\le snr.
 \EE
Also, by definition, the MMSE $\rm NLD_{DEC}$ in \eqref{Eqn:NLD_MMSE_coded} should achieve an MSE lower than that of a symbol-by-symbol detector, i.e.,
\BE\label{Eqn:coding_gain}
\omega_{{\mathcal{C}}}(\rho)< \omega_{\cal S}(\rho),  \;\;\;  {\rm for} \;\; \rho \geq 0.\vspace{-0.1cm}
\EE\ES
Combining \eqref{Eqn:error-free} and \eqref{Eqn:coding_gain}, we obtain a necessary and sufficient condition for AMP to achieve error-free detection:
\BS\label{Eqn:upper_bound}\BE
\omega_{\mathcal{C}}({\rho}) < \omega_{\mathcal{C}}^*({\rho}),\;\;\; {\rm for} \;\; 0 \leq {\rho } \leq snr,
\EE
where
\BE \label{Eqn:w_star}
 \omega_{\mathcal{C}}^*({\rho})=  
   \min\big\{ \omega_{\cal S}(\rho),\; \phi^{-1}({\rho }) \big\}, \quad  {\rm for} \;\; 0 \le {\rho }\le  snr.
\EE\ES

\begin{proposition}\label{The:area_LRMS}
Suppose that Assumption \ref{Pro:SCP} holds.  Then the constrained capacity of an LRMS with the given $\mathcal{\bf{S}}$ is   
\BS\label{Eqn:dis_cap_new2}\BE
C=A_{\omega_{\mathcal{C}}^*},
\EE 
where $A_{\omega_{\mathcal{C}}^*}$  is the area covered by $\omega_{\mathcal{C}}^*$, i.e.,  
\begin{align}
A_{\omega_{\mathcal{C}}^*}&\equiv\int_{0}^{snr} \omega_{\mathcal{C}}^*({\rho}) d\rho\\ 
& =    \beta^{-1}\big[\rho^{*}/snr\!-\!\log(\rho^{*}/snr)\!-\!1\big] \!+\!\! \int_0^{\rho^{*}} \!\!\! \omega(\rho) d \rho.
\end{align} \ES
\end{proposition}
\begin{IEEEproof} 
See APPENDIX \ref{APP:Consistency}. 
\end{IEEEproof}

Combining \eqref{Eqn:R_mmse}, \eqref{Eqn:upper_bound} and \eqref{Eqn:dis_cap_new2}, we obtain the capacity optimality of AMP below.
\begin{theorem}[Capacity Optimality]\label{The:cap_opt}
Assume that Assumptions \ref{Pro:SCP} and \ref{Pro:SE_new} hold and AMP converges to $v=0$. Then, 
\BE
R_{\cal C}\to C,
\EE 
if $\omega_{\mathcal{C}}(\rho)\to \omega_{\mathcal{C}}^*(\rho)$ in $[0, snr]$.
\end{theorem}
 
\begin{figure}[t]
  \centering
  \includegraphics[width=7cm]{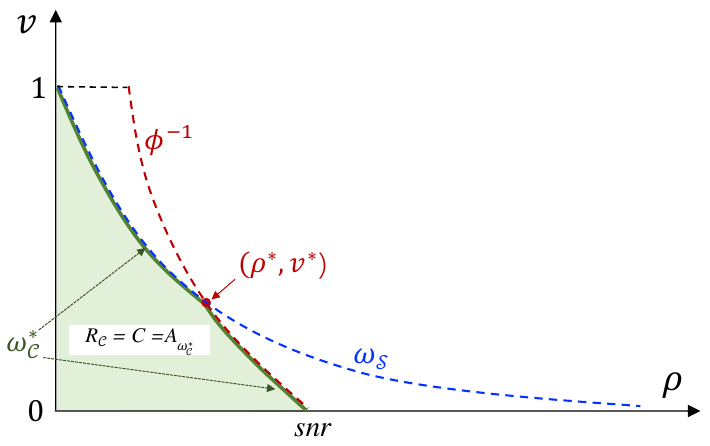}\\ 
  \caption{Graphical illustration of the capacity, the maximum achievable rate of AMP and the optimal transfer function of decoder. The maximum achievable rate of AMP equals to the capacity, which is the area covered by  $\omega_{\mathcal{C}}^*$. }\label{Fig:Area_A}
\end{figure}

Fig. \ref{Fig:Area_A} gives a graphical illustration of Theorem \ref{The:cap_opt}. The following are some notes on Theorem~\ref{The:cap_opt}.  
\begin{itemize}
    \item Theorem \ref{The:cap_opt} is based on a matching condition:
\BE\label{Eqn:C2Copt}
\omega_{\mathcal{C}}(\rho)\to \omega_{\mathcal{C}}^*(\rho).
\EE
A proof for the existence of a code achieving \eqref{Eqn:C2Copt} can be found in Appendix \ref{APP:Gau_SCM} for Gaussian signaling. For other signaling, the existence of such a code is a conjecture only. We will discuss techniques to approximately achieve \eqref{Eqn:C2Copt} for QPSK modulations in \ref{Sec:ldpc_opt}. 
\item The situations for multi-ary modulations are more complicated. Various techniques have been developed to match the extrinsic information transfer (EXIT) functions of two local processors in conventional Turbo receivers \cite{Hanzo09,Hanzo10,Matsumoto14}. These methods can be borrowed to achieve \eqref{Eqn:C2Copt} approximately. Detailed discussions on the multi-ary systems are beyond the scope of this paper.  
\item Interestingly, the discussions above provide an alternative proof for the constraint capacity of an LRMS. The key is to prove $A_{\omega_{\mathcal{C}}^*}=C$ without prompting the result in \cite{Barbier2017arxiv, Barbier2016conf, Reeves_TIT2019}.  This is indeed possible using the properties of AMP directly. The details can be found in Appendix \ref{APP:the_dis_cap}. 
\end{itemize}

\begin{figure}[t] 
  \centering
  \includegraphics[width=6.75cm]{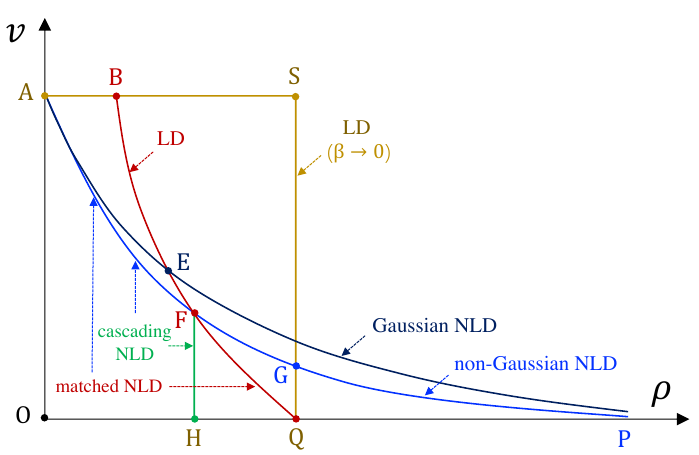}\\
  \caption{Interpretations of SINR-variance transfer charts and areas of AMP. Curve BQ represents LD $\phi^{-1}$. When $\beta \to 0$, it moves to the ``interference-free" SQ. AEP: un-coded Gaussian NLD $\omega_{\mr{Gau}}$; AFP: un-coded non-Gaussian NLD $\omega_{\mathcal{S}}$; AFQ: matched decoder $\omega_{{\mathcal{C}}}^*$; AFH: cascading decoder.}\label{Fig:area_expl} 
\end{figure}

{\emph{Area Properties:}}  
Based on the above discussions, we can obtain some interesting area properties as illustrated in Fig. \ref{Fig:area_expl}.  
\begin{enumerate}[(i)]
\item Area $A_{\mathrm{AGPO}}$ gives the entropy of the constellation, e.g. $\log|\mathcal{S}|$, as discussed in \ref{Sec:SISO_area}, which is maximum achievable rate in the noiseless case.
\item Area $A_{\mathrm{AGQO}}$ gives the capacity of SISO-AWGN channel as discussed in \ref{Sec:SISO_area} (see Fig. \ref{Fig:AWGN_area}).
\item Area $A_{\mathrm{GPQ}}$ gives the rate loss after SISO coding to combat the channel noise. As SNR goes to infinity, point Q moves right to infinity (e.g. point P) and the SISO code rate approaches the maximum $\log|\mathcal{S}|$.

\item Area $ A_{\mathrm{AFQO}}$ gives the capacity $C$ of an LRMS and also the achievable rate of AMP.
\item Area $A_{\mathrm{AFHO}}$ gives the achievable rate of a receiver with an AMP detector cascaded by a decoder \cite{Guo2005random, Tanaka2002}. There is no iteration between the two. In this case, AMP achieves the MMSE by treating the codeword as an IID sequence. However, the overall algorithm is not capacity optimal (see \ref{Sec:Comparisons}).\label{Area:casd}
\item Area $A_{\mathrm{FQH}}$ gives the rate loss for the cascading receiver in \eqref{Area:casd}.

\item Area $A_{\mathrm{AEF}}$ gives the shaping gain of Gaussian signaling. Curve AEP denotes the un-coded Gaussian NLD for Gaussian signaling.
\item  Area $A_{\mathrm{FGQ}}$ gives the capacity gap of LRMS and parallel SISO channels. When $\beta\to0$, $ \lim_{\beta\to0} \bf{A}^{\rm H}\bf{A}=\bf{I}$, indicating that the capacity of an LRMS will converge to that of a set of parallel SISO channels in the limiting case. In this case, the cross-symbol interference disappears, so B$\to$S and areas $A_{\mathrm{FGQ}}$ and $A_{\mathrm{FQH}}$ become negligible. Then, the separate detection and decoding strategy becomes optimal. 
\end{enumerate}

{\subsection{Gaussian Signaling}\label{Sec:Gau_Sig}
We now study a special case of \ref{Sec:area_AMP} when $\bf{x}$ is Gaussian. We show that in this case both Assumptions \ref{Pro:SCP} and \ref{Pro:SE_new} asymptotically hold. In addition, the results in \ref{Sec:area_AMP} have simpler derivations as well as closed-form expressions. 

For Gaussian signaling, $ \omega(\rho)=\omega_{\mr{Gau}}(\rho)= {1}/(1+\rho)$ \cite{Guo2005random}. Thus we can rewrite \eqref{Eqn:w_star} as
\BE
  \omega_{\mathcal{C}-{\mr{Gau}}}^*(\rho)\!=\!\min\{1/(\rho+1),\; \phi^{-1}({\rho })\},\;\; 0 \!\leq \!{\rho } \!\leq\! snr.
\EE
Then we have the following results for an LRMS with Gaussian  signaling. 
\begin{itemize}
    \item {\emph{Unique Fixed Point:}} {Assumption \ref{Pro:SCP}  holds}. That is, equation $\omega_{\mr{Gau}}(\rho)=\phi^{-1}(\rho)$ has a unique positive solution:
    \BE\label{Eqn:Gau_rho}
    \!\!\rho^{*}_{\mr{Gau}}\!=\!\dfrac{(1\!-\!\beta)snr\!-\!1\!+\!\sqrt{[(1\!-\!\beta)snr\!-\!1]^2\!+\!4snr}}{2}.
    \EE  

    \item {\emph{Area Property:}}  Let the area covered by  $\omega_{\!\mathcal{C}-{\mr{Gau}}}^*$ be 
    \BS\label{Eqn:Gau_area}\begin{align}
        &\hspace{-2.5mm} A_{\omega_{\mathcal{C}-{\mr{Gau}}}^*}  \equiv\int_{0}^{snr} \!\!\!\omega_{\mathcal{C}-{\mr{Gau}}}^*({\rho}) d\rho \\
        & \hspace{-2.5mm} = \! \beta^{-1}\!\log(1\!+\!\beta \,snr\, v^{*}_{\mr{Gau}}) \!-\! \log(v^{*}_{\mr{Gau}}) + v^{*}_{\mr{Gau}}\!\!-\!1, 
    \end{align}\ES
    where $v^{*}_{\mr{Gau}}\!=\!\phi^{-1}(\rho^{*}_{\mr{Gau}})$. Then area $A_{\omega_{\mathcal{C}-{\mr{Gau}}}^*}$ equals to the Gaussian capacity of an LRMS, i.e.,
     \BE\label{Eqn:dis_cap_Gau}
        A_{\omega_{\mathcal{C}-{\mr{Gau}}}^*}= C_{\mr{Gau}}.
    \EE  
     \begin{IEEEproof}
        See APPENDIX \ref{APP:Gau_Instance}.
    \end{IEEEproof}

    \item {\emph{Code Existence:}} For Gaussian signaling, there exists an $n$-layer superposition coded modulation  (SCM) code with rate $R_n$ and transfer function $\{\omega_{\mathcal{C}_n}(\rho)< \omega_{{\mathcal{C}-{\mr{Gau}}}}^*(\rho), \forall \rho\geq 0\}$, and as $n\to \infty$,
    \BE
        R_n \to A_{\omega_{\mathcal{C}-{\mr{Gau}}}^*}.
    \EE
    \begin{IEEEproof}
        See Appendix \ref{APP:Gau_SCM}.
    \end{IEEEproof} 
    That is, for Gaussian signaling, there exists an SCM code that asymptotically matches with $\omega_{\mathcal{C}-{\mr{Gau}}}^*$ and its rate $\to A_{\omega_{\mathcal{C}-{\mr{Gau}}}^*}$. 
    \item {\emph{Capacity Optimality:}} Assume that Assumption \ref{Pro:SE_new} holds, AMP achieves the Gaussian capacity when $\omega_{\mathcal{C}}=\omega_{\mathcal{C}-{\mr{Gau}}}^*$:
        \BE
            R_{\cal C}=C_{\mr{Gau}}.
        \EE  
\end{itemize}} \vspace{-5mm}

\subsection{Comparisons with Alternative Algorithms} \label{Sec:Comparisons}
 
{\emph{1) Comparison with Turbo-LMMSE:}} 
It is proved in \cite{Lei20161b,Yuan2014} that Turbo-LMMSE is capacity achieving for Gaussian signaling. In the following, we show that Turbo-LMMSE is sub-optimal for non-Gaussian signaling.

{The main difference between AMP and Turbo-LMMSE is as follows. To avoid the correlation problem in the iterative process, Turbo-LMMSE uses extrinsic local processors (e.g. an extrinsic LD and an extrinsic decoder), while AMP uses an ``Onsager"-term.}

Assume that the transfer functions of the detector and the decoder in Turbo-LMMSE are matched. The achievable rate of Turbo-LMMSE is given in \cite{Yuan2014} 
\BE
R_{\mr{LMMSE}}=\log|\mathcal{S}|-\int_0^{+\infty} \omega_\mathcal{S}\big(\rho+\phi\big(\omega_\mathcal{S}(\rho)\big)\big) d\rho.
\EE

\begin{figure}[t] 
  \centering
  \includegraphics[width=8.8cm]{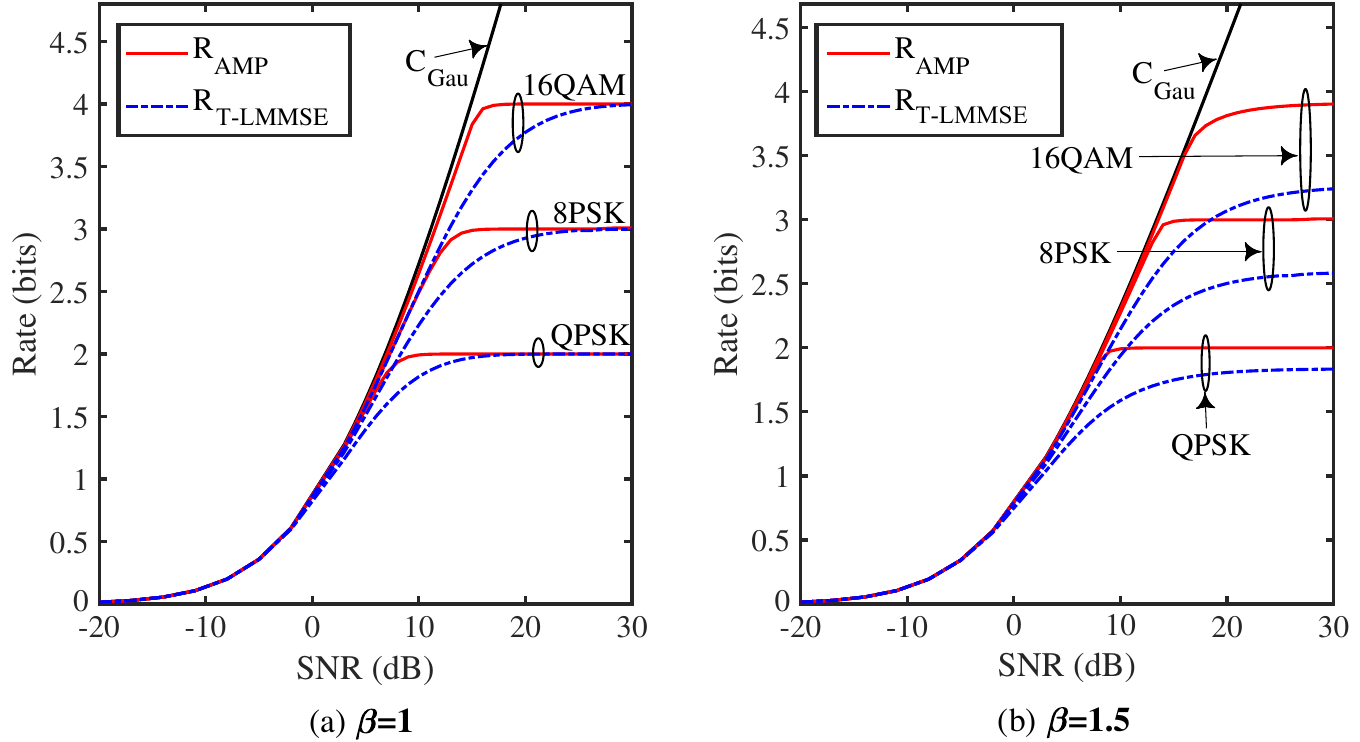}\\ 
  \caption{Comparison between the capacity and the achievable rates of AMP and Turbo-LMMSE of an LRMS with $\beta=N/M=\{1, 1.5\}$, where  $C_{\rm Gau}$ denotes the Gaussian capacity and also the achievable rates of AMP and Turbo-LMMSE with Gaussian signaling, $R_{\mr{AMP}}$ and $R_{\mr{T-LMMSE}}$ respectively denote the achievable rates of AMP and Turbo-LMMSE with QPSK, 16QAM and 8PSK modulations.}\label{Fig:Rate_AMP_Turbo} 
\end{figure}
Fig. \ref{Fig:Rate_AMP_Turbo} shows the capacity and the achievable rates of AMP and Turbo-LMMSE. The capacity for Gaussian signaling is achieved by both AMP and Turbo-LMMSE. For QPSK, 8PSK and 16QAM modulation, the achievable rate of AMP equals to capacity when {Assumption \ref{Pro:SCP} holds}, while Turbo-LMMSE always has rate loss. Similar results can be obtained for other non-Gaussian signaling. In addition, the gap between AMP and Turbo-LMMSE increases with $\beta$. This gap $\to0$ when $\beta\to0$. {The reason why Turbo-LMMSE has performance loss is that extrinsic update leads to performance loss for non-Gaussian signal processing, which was first pointed out in \cite{MaTWC}. For more details, please refer to \cite{MaTWC}.}

{\emph{2) Comparison with Cascading AMP and Decoding:}}  
We define a cascading AMP and decoding (AMP-DEC) scheme \cite{Guo2005random, Tanaka2002} as follows. We run AMP until it converges. The result is used by decoder. There is no iteration between AMP and the decoder. The area $A_{\mr{AFHO}}$ in Fig. \ref{Fig:area_expl} shows the achievable rate of AMP-DEC, i.e.,
\BE\label{Eqn:dis_ach}
R_{\mr{AMP-DEC}}=  A_{\mr{AFHO}}= \int_0^{{{\rho}}^{*}} \!\!\! \omega_\mathcal{S}(\rho) d \rho.
\EE
For Gaussian signaling, $\omega_{\mr{Gau}}(\rho)= {1}/{1+\rho}$. Hence,
\BE\label{Eqn:Gau_ach}
R_{\mr{AMP-DEC}}  = \log(1+\rho^*_{\mr{Gau}}),
\EE
where $\rho^{*}_{\!\mr{Gau}}\!=\!0.5\left[\!{(1\!-\!\beta)snr\!-\!1\!+\! \sqrt{[(1\!-\!\beta)snr\!-\!1]^2\!+\!4snr}} \right]$  (see \eqref{Eqn:Gau_rho}). If $\beta>1$, when $snr\to \infty$, we have 
\BE
\rho^*_{\mr{Gau}}\to (\beta-1)^{-1},
\EE
and 
\BE
R_{\mr{AMP-DEC}}  \to -\log(1-\beta^{-1}). 
\EE
That is, the achievable rate of AMP-DEC converges to a finite value, and it goes to zero as $\beta \to \infty$. This is very different from the Gaussian system capacity that $C\to \infty$ as $snr\to \infty$.

\begin{figure}[t] 
  \centering
  \includegraphics[width=8.8cm]{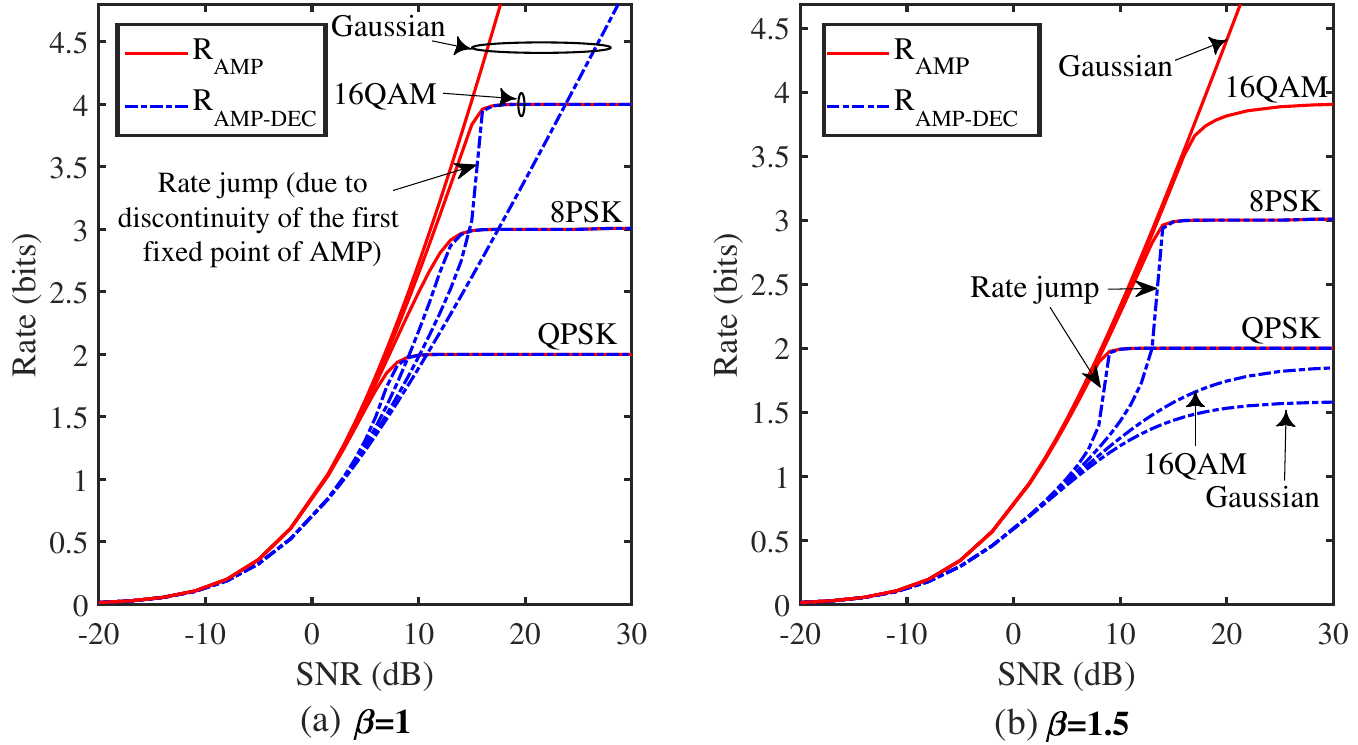}\\ 
  \caption{Comparison between the capacity, the achievable rates of AMP, and separate optimal MMSE detection and ideal SISO decoding in \cite{Guo2005random, Tanaka2002} with $\beta=N/M=\{1, 1.5\}$, where $C_{\rm Gau}$ denotes the Gaussian capacity and also the achievable rates of AMP with Gaussian signaling, $R_{\mr{AMP}}$ and $R_{\mr{AMP-DEC}}$ respectively denote the achievable rates of AMP and ``cascading AMP and decoding" scheme with QPSK, 16QAM and 8PSK modulations. }\label{Fig:Rate_AMP_AWGN} 
\end{figure}

Fig. \ref{Fig:Rate_AMP_AWGN} compares AMP and AMP-DEC. For QPSK, 8PSK and 16QAM modulations, the achievable rate of AMP-DEC is lower than that of AMP. This gap increases with $\beta$, but is negligible if $\beta$ is small (e.g. $\beta<0.5$ based on our experimental findings). Furthermore, different from the rate of AMP that always increases with the size of constellation, the rate of AMP-DEC decreases with the increasing of the constellation size for large $\beta$.

  \begin{figure}[t]
  \centering
  \includegraphics[width=6.5cm]{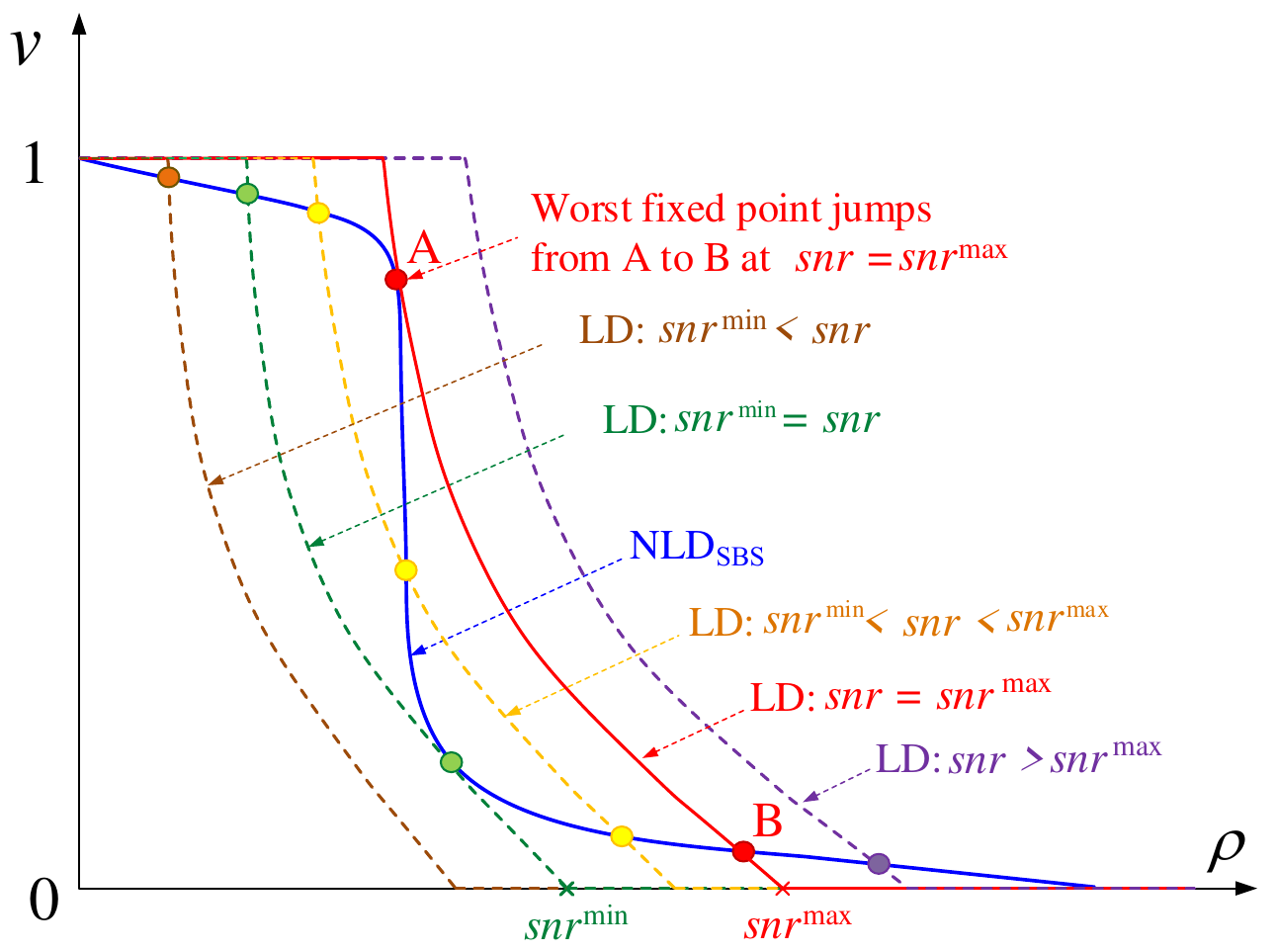}\\ 
  \caption{Multiple fixed points illustration of AMP with QPSK modulation, $\beta=N/M=2$, $snr^{\mr{min}}=9.05$ dB, $snr^{\mr{max}}= 15.77$ dB. { Assumption \ref{Pro:SCP} holds} when $snr<snr^{\mr{min}}$ or  $snr>snr^{\mr{max}}$; Three fixed points for $snr^{\mr{min}}<snr<snr^{\mr{max}}$. Besides, the first fixed point jumps from Point A to Point B at $snr=snr^{\mr{max}}$. ``semilogy"-plot is used to see the cross points clearly.}\label{Fig: Multi_Fixed_Points} 
\end{figure} 

As shown Fig. \ref{Fig:Rate_AMP_AWGN}, the achievable rate of AMP-DEC  jumps at certain $snr$ values. This happens when the number of fixed points changes. Fig. \ref{Fig: Multi_Fixed_Points} illustrates this phenomenon for $\beta=2$. For $snr=snr^{\mr{max}}$, the number of fixed points is 3 at the left vicinity and 1 at the right vicinity. While increasing $snr$, the worst fixed point jumps from Point A to Point B at $snr=snr^{\mr{max}}$, resulting in rate jump of AMP-DEC in Fig. \ref{Fig:Rate_AMP_AWGN}.

\begin{figure}[t]  
  \centering
  \includegraphics[width=5cm]{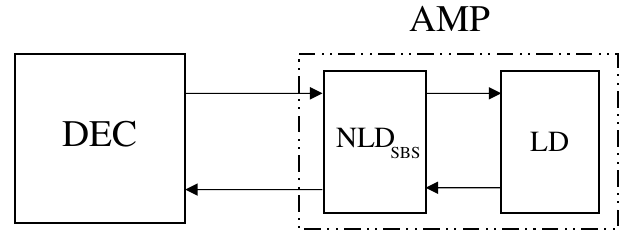}\\ 
  \caption{AMP with internal iteration, in which ``DEC'' denotes decoder, and AMP is discussed in \ref{Sec:AMP_uncoded} with SBS-NLD being the demodulation for the constellation constraint $\cal S$. }\label{Fig:Conv_AMP} 
\end{figure}

{ {\emph{3) Comparison with AMP with Internal Iteration:}} Fig. \ref{Fig:Conv_AMP} illustrates an alternative for AMP. It involves the iteration of a decoder (DEC) module and an AMP module. In each global iteration, there are multiple internal iteration within the AMP module. The scheme in Fig. \ref{Fig:Conv_AMP} may have the advantage of low cost if the overall complexity is dominated by that of DEC. However, if the decoding complexity (e.g. sum-product LDPC decoding) is lower than that of AMP, the scheme in Fig. \ref{Fig:Conv_AMP} may have higher complexity due to the internal iteration. Since there is no closed-form transfer function for the AMP module, it makes the achievable rate analysis and optimization more difficult. We conjecture that the scheme in Fig. \ref{Fig:Conv_AMP} has the same overall performance of AMP without internal iteration. However, we do not have a proof of this conjecture.} 

\section{LDPC Code Design and Simulation Results}\label{Sec:SIM}
This section discusses matching techniques using optimized LDPC codes and QPSK modulation. Simulation results will be provided. 

{\subsection{LDPC Code Optimization for AMP}\label{Sec:ldpc_opt} 
According to the code design principle in Section \ref{Sec:Cap_Opt}, the object of  code optimization is to design an code with an \emph{a-posterior} variance transfer function $v=\omega_{\cal C}(\rho)$ satisfying the following matching condition:
\BE\label{Eqn:giv_curv}
v=\omega_{\cal C} (\rho)= \omega_{\cal C}^*(\rho), 0\le \rho\le snr,
\EE
with a coding rate equal to the capacity.

The extrinsic information transfer (EXIT) chart matching techniques \cite{Brink2001,Brink2004,Yuan2008Low} can be used for this purpose by choosing optimized degree distributions. The discussions below follow \cite[Appendix 5G]{Yuan2008Low} to design irregular LDPC codes. The difference is that mutual information is used instead of log likelihood ratio (LLR) in tracking the evolution process.

According to \cite{Brink2004}, the decoder characteristic for an LDPC code can be computed as 
\BS\begin{align}
I_{\rm E,V}\! &=\!\! \textstyle\sum\limits_{\!i=1}^{\!d_{v,\max}}\!\lambda_i \cdot J \! \left( \sqrt{(i\!-\!1)\left[ J^{-1}\left(I_{\rm E,C}\right) \right]^2\! + \!4\rho} \right), \label{eq:variable_evo} \\
I_{\rm E,C}\! &=\! 1 \!-\!\!\textstyle\sum\limits_{j=1}^{d_{c,\max}}\eta_j \cdot J\left( \sqrt{j\!-\!1} \cdot J^{-1}\left(1\!-\!I_{\rm E,V} \right) \right),  \label{eq:check_evo}
\end{align}\ES
 where 
 \begin{itemize}
     \item $I_{\rm E,V}$ (resp., $I_{\rm E,C}$) is the extrinsic information from variable node (resp., check node) to check node (resp., variable node),
     \item $d_{v,\max}$ (resp., $d_{c,\max}$) is the maximum variable node (resp., check node) degree,
     \item $\lambda_i$ (resp., $\eta_i$) is the fraction of edges in the bipartite graph of the LDPC code connected to variable nodes (resp., check node) with degree $i$,     
     \item the function $J(\cdot)$ is
     \begin{equation}\label{eq:Jfunc}
        J(\sigma_{ch}) \!= \!1\! -\!\! \int_{-\infty}^{\infty}\!\frac{e^{-\frac{(y-\sigma_{ch}^2\!/\!2)^2}{2\sigma_{ch}^2}}}{\sqrt{2\pi\sigma_{ch}^2}} \log_2\left( 1\!+\!e^{-y} \right) dy,
    \end{equation} 
    and $J^{-1}(\cdot)$ is the inverse of $J(\cdot)$,
    \item $\rho$ is the decoder input SNR. 
 \end{itemize}
Substituting \eqref{eq:check_evo} into \eqref{eq:variable_evo}, we can charaterize the LDPC code by one single variable $I_{\rm E,V}$ using  
\BE\label{eq:evo}
I_{\rm E,V} \!=  \!\!\!\sum\limits_{i=1}^{d_{v,\max}}\!\!\lambda_i  J \! \left(\! \sqrt{(i\!-\!1)\!\left[ J^{-1}\big( I_{\rm E,C}(I_{\rm E,V}) \big)   \right]^2 + 4\rho}\right).
\EE

To satisfy \eqref{Eqn:giv_curv}, the converged extrinsic information $I_{\rm E,V}$ of the LDPC decoder, denoted as $I_{E,V, \mr{fin}}(\rho)$, should lead to an output \emph{a posterior}  variance $v=\omega_{\cal C}(\rho)$, i.e., given $\rho$, $I_{E,V, \mr{fin}}(\rho)$ should satisfy the following equation.
\BE\label{Eqn:I_V}
\sum\limits_{\!i=1}^{\!d_{v,\max}}\!\!\Lambda_i   \omega_{\mr{QPSK}} \left( \frac{i [J^{-1}(I_{E,C,\mr{fin}}) ]^2\!+\!4\rho}{4} \right) = \omega_{\cal C}(\rho),
\EE
where
\begin{itemize}
    \item $\Lambda_i$ is the fraction of variable node of degree $i$ and is computed as 
    \BS\begin{equation}
        \Lambda_i = \left. \lambda_i / i \middle/\; \textstyle\sum\limits_{i=1}^{d_{v,\max}} \lambda_i / i \right.,
    \end{equation}
    \item $\omega_{\mr{QPSK}}(\cdot)$ is the MMSE function of QPSK demodulation given in \eqref{Eqn:QPSK_MMSE},
    \item $I_{E,C,\mr{fin}}$ is the converged check node to variable node message and is computed as
    \begin{equation}
    I_{E,C,\mathrm{fin}}\! = \! 1 \!-\!\! \textstyle\sum\limits_{\!j=1}^{\!d_{\!c,\max}}\! \eta_j  J\left( \sqrt{j\!-\!1}   J^{-1}\left(1\!-\!I_{\rm E,V, fin} \right) \right).
    \end{equation} \ES
\end{itemize}
The optimization problem can be formulated to maximize the code rate under the constraint \eqref{Eqn:I_V}:
\begin{align}
    \max\limits_{\{\lambda_i\}}  & \textstyle\sum\limits_{i=1}^{d_{v,\max}}  {\lambda_i}/{i} \qquad 
\mbox{s.t.~} \textstyle\sum\limits_{i=1}^{d_{v,\max}} \lambda_i = 1,\;\; \{\lambda_i\}\in \Xi.  \label{eq:LDPC_opt}   
\end{align} 
where $\Xi$ is defined as the constraint:
\BS\BE   
 \sum\limits_{i=1}^{d_{v,\max}}\!\lambda_i  J \! \left( \!\!\sqrt{(i\!-\!1)\!\left[ J^{-1}\big( I_{\rm E,C}(I_{\rm E,V}) \big) \!\right]^2 \!\!+ 4\rho} \right)\! >\! I_{\rm E,V}, 
\EE 
for $\forall 0 < \rho < \infty $ and $I_{\rm E,V, ini}(\rho) \leq I_{\rm E,V} \leq I_{\rm  E,V,\mathrm{fin}}(\rho)$, and $I_{\rm E,V, ini}(\rho)$ is the initial extrinsic information given by the channel: 
\begin{align}
  I_{\rm E,V\!, ini}(\rho)\! =\! \!  \!\!\textstyle\sum\limits_{\!i=1}^{\!d_{v,\max}}\!\!  \lambda_i  J \! \left(\! \!  \sqrt{(i\!-\!1)\left[ J^{-1}\left(0\right) \right]^2 \!+\! 4\rho} \right) \!\! =\! J ( 2\sqrt{\rho} ).  \label{eq:ini_IEV}
\end{align} \ES

The optimization problem in \eqref{eq:LDPC_opt} is non-convex. However, given $\{\eta_i, i=1,2,\dots,d_{c,\mr{max}}\}$, the problem in \eqref{eq:LDPC_opt} can be solved using a standard linear programming technique. In this paper, an iterative way to optimize $\{\lambda_i, i=1,2,\dots,d_{v,\mr{max}}\}$ with $\{\eta_i, i=1,2,\dots,d_{c,\mr{max}}\}$ fixed is used (see Algorithm \ref{alorithm:LDPCopt}). 
In Algorithm \ref{alorithm:LDPCopt}, degree distributions are denoted as 
\BE
\lambda(x) = \textstyle\sum\limits_{\!i=1}^{\!d_{v,\max}}\!\lambda_i x^{i-1} \;\; \mr{and} \;\; \eta(x) = \textstyle\sum\limits_{\!i=1}^{\!d_{c,\max}}\!\eta_i x^{i-1}.
\EE

We observed that a maximum trial $T=5$ and a threshold $\epsilon=10^{-3}$ are good choice for Algorithm \ref{alorithm:LDPCopt}.  Also, we force the degree-1 fraction $\lambda_1=0$. For results given in Table~\ref{Opt_degree1}, Algorithm \ref{alorithm:LDPCopt} is repeated by manually tuning  check edge distribution $\eta(x)$ and maximum variable degree $d_{v,\max}$ until a matching code is found. For the LDPC code optimization for Turbo-LMMSE, we only need to change \eqref{Eqn:I_V} into to 
\BE\label{Eqn:I_V2}
\textstyle\sum\limits_{\!i=1}^{\!d_{v,\max}}\!\Lambda_i \cdot \omega_{\mr{QPSK}} \left( \frac{i\cdot[J^{-1}(I_{E,C,\mr{fin}}) ]^2}{4} \right) = \omega^{\mr{ext}}_{\cal C}(\rho),
\EE
where $\omega^{\mr{ext}}_{\cal C}(\rho)$ is matched to the Turbo-LMMSE LD transfer function.

\begin{algorithm}[b!]
\caption{Algorithm for LDPC Code Optimization}
 \begin{algorithmic}[1]\label{alorithm:LDPCopt}
 \renewcommand{\algorithmicrequire}{\textbf{Input:}}
 \renewcommand{\algorithmicensure}{\textbf{Output:}}
 \REQUIRE Target decoder transfer $v = \omega_{\cal C}(\rho)$, check edge distribution $\eta(x)$, maximum trial $T$, Threshold $\epsilon$ and maximum variable degree $d_{v,\max}$.\vspace{2mm}
 \ENSURE The optimized variable edge distribution $\lambda^{(T)}(x)$.\vspace{2mm}
  \STATE Initialize $\lambda^{(0)}(x) = x$.\vspace{2mm}
  \FOR {$t = 1$ to $T$}\vspace{2mm}
  \STATE Solve (\ref{eq:LDPC_opt}) by linear programming to obtain $\lambda^{(t)}(x)$, where $I_{E,V,\mbox{fin}}(\rho)$ in (\ref{eq:LDPC_opt}) is obtained by solving \eqref{Eqn:I_V} using $\lambda^{(t-1)}(x)$.\vspace{2mm} 
  \IF {1-$\frac{\sum\limits_{i=1}^{d_{v,\max}}\lambda_i^{(t)}\lambda_i^{(t-1)}}{\sqrt{\left(\sum\limits_{i=1}^{d_{v,\max}}(\lambda_i^{(t)})^2\right) \left(\sum\limits_{i=1}^{d_{v,\max}}(\lambda_i^{(t-1)})^2\right)}} \leq \epsilon$}\vspace{3mm}
  \STATE $\lambda^{(T)}(x) = \lambda^{(t)}(x)$.\vspace{2mm}
  \RETURN $\lambda^{(T)}(x)$.\vspace{2mm}
  \ENDIF
  \ENDFOR
 \RETURN $\lambda^{(T)}(x)$.
 \end{algorithmic}
\end{algorithm} 

With the optimized degree distributions $\eta(x)$ and $\lambda^{(T)}(x)$, a parity-check matrix for simulation is generated as follows:
\begin{itemize}
    \item Set a code length to $10^5$.
    \item Randomly generate a parity-check matrix of columns equal to the code length according to the optimized degree distributions $\eta(x)$ and $\lambda^{(T)}(x)$.
    \item Remove cycle-4 loops in the generated parity-check matrix by removing one edge in every founded cycle-4 loop in the parity-check matrix.
\end{itemize}
Removing cycle-4 loops using the above method will slightly change the degree distributions. However, for a code length of $10^5$, this change has negligible effect on overall performance. 

The high error floors in Fig. \ref{Fig:BER_AMP} are due to the fixed points of $\phi^{-1}(\cdot)$ and $\omega_{\cal C}(\rho)$. More research effort is still required to improve this problem.}

{ {\emph{Related Works:}} Recently, LDPC codes  are optimized to support much higher sum spectral efficiency and user loads for linear systems \cite{Yuan2008Low, Chung2001, YHu2018}. Based on the EXIT  analysis~\cite{Brink2001, Brink2004}, a LDPC code is constructed to obtain a near capacity performance \cite{XWang20182,XWang2020}. To support massive users, an irregular repeat-accumulate (IRA) code is optimized in \cite{SongTVT, Song-MaxSum}. More recently, a Turbo-LMMSE receiver with an optimized IRA code approaches the capacity for various of system loads \cite{YC2018TWC}. However, the results in \cite{YC2018TWC} mainly focused on low-rate coding schemes (e.g. $R_{\cal C} =$ 0.1 or 0.2). In these cases, Turbo-LMMSE is near-optimal (see the region of  $R_{\cal C} \le0.5$ in Fig. \ref{Fig:Rate_AMP_Turbo}). In this paper, we will show that AMP performs much better than Turbo-LMMSE in high transmission rate.

For higher order modulation, the design of a curve-matching code is more complicated. Irregular bit-interleaved coded modulation (Ir-BICM) was developed for high-order modulation \cite{Hanzo09,Hanzo10,Matsumoto14}. These methods can be borrowed to design curve-matching codes for high-order modulations. Detailed discussions on the high-order modulations are beyond the scope of this paper. We leave it as our future work.}

\subsection{BER Comparison with AWGN Irregular LDPC Code and Regular LDPC Code}\label{Sec:BER_per}

\begin{figure*}[t]
  \centering
  \includegraphics[width=15cm]{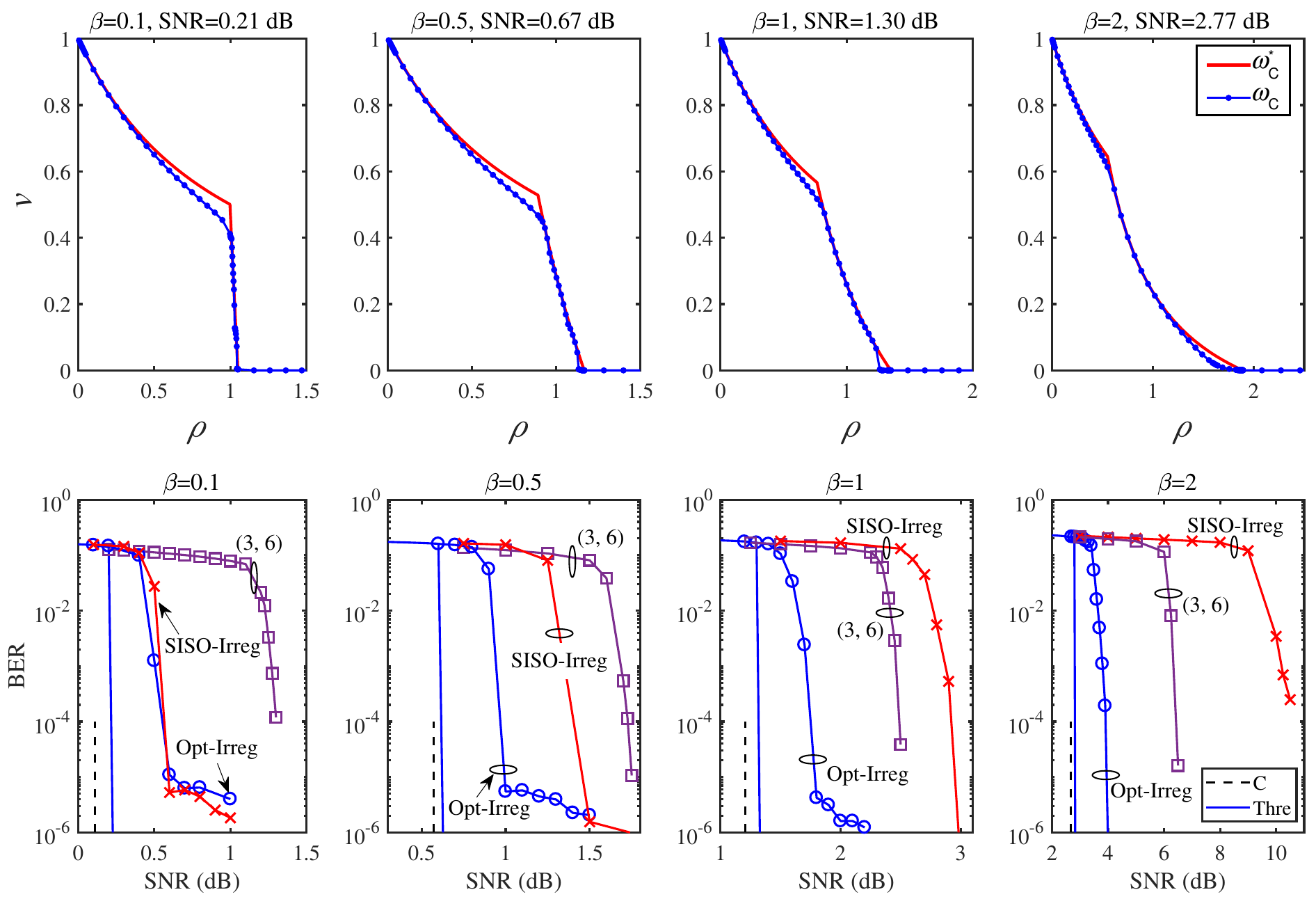}\\
  \caption{Transfer function matching and BER performances of AMP, where ``$\omega_{\mathcal{C}}^*$''  denotes the fully matched transfer function of AMP (target), ``$\omega_{\mathcal{C}}$'' the optimized transfer function of decoder of AMP,
   $C$ the capacity limit, ``Thre''  the BER threshold, ``Opt-Irreg'' the BER of AMP-optimized irregular LDPC codes, ``SISO-Irreg'' the BER with SISO-optimized irregular LDPC codes, ``(3, 6)''  the BER of AMP with regular (3, 6) LDPC code. Code length = $10^5$, code rate $\approx$ 0.5, QPSK modulation, iterations = $200\sim700$, and $\beta=N/M=\{0.1,0.5,1,2\}$. For more details, refer to Table \ref{Opt_degree1}.}\label{Fig:BER_AMP} 
\end{figure*}
\begin{figure*}[t] 
  \centering
  \includegraphics[width=15cm]{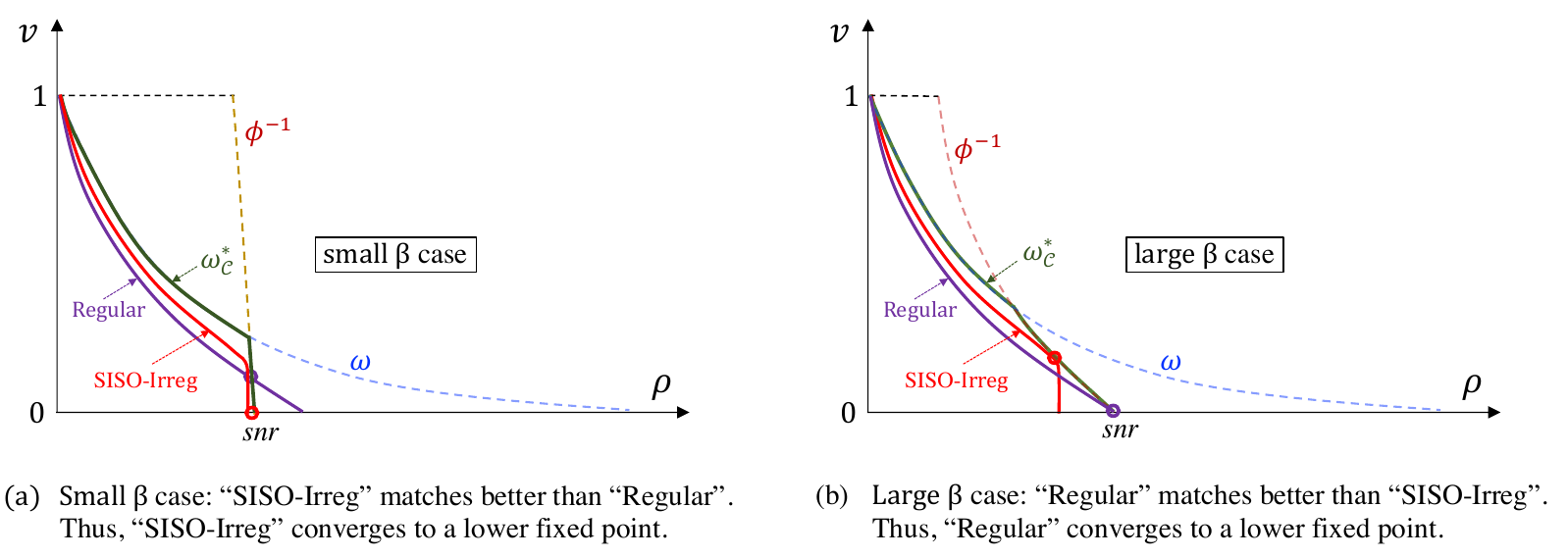}\\ 
  \caption{Graphical illustration of AMP with the regular $(3,6)$ LDPC code and the SISO-Irreg LDPC code.}\label{Fig:beta_match} 
\end{figure*}

Fig. \ref{Fig:BER_AMP} provides the BER simulations for an LRMS, in which $\bf{x}$ is generated using optimized irregular LDPC codes \cite{Yuan2008Low, Chung2001} {with code length $=10^5$}. The AMP (see Fig. \ref{Fig:model_coded}) for an optimized LDPC coding (see \ref{Sec:ldpc_opt}) LRMS is denoted as ``Opt-Irreg''. The APP decoder is implemented using a standard sum-product decoder. The channel loads are $\beta=\{0.1, 0.5, 1, 2\}$ with $(N, M)\!=\!(250, 2500), (250, 500), (500, 500)$ and $(500, 250)$, respectively. The corresponding optimized code parameters are given in Table~\ref{Opt_degree1}, which illustrates that these decoding thresholds are very close (about $0.1$ dB$\sim$$0.2$ dB away) to the Shannon limits.

\newcommand{\tabincell}[2]{\begin{tabular}{@{}#1@{}}#2\end{tabular}}
\renewcommand\arraystretch{1.15}
\begin{table*}[t] \footnotesize
\caption{Optimized Irregular LDPC Codes for AMP and Turbo-LMMSE under QPSK Modulation}\label{Opt_degree1} 
\centering\setlength{\tabcolsep}{1mm}{
\begin{tabular}{|c||c|c|c|c|c|c|}
\hline
Methods & \multicolumn{5}{c|}{AMP}   & Turbo-LMMSE\\
\hline
& \multicolumn{5}{c|}{\vspace{-0.35cm}}   &\\
\hline
$\it{\beta}$ & {$0.1$} & {$0.5$}& {$1$} & $1.5$ & {$2$}& $1.5$ \\
\hline
$\textit{$N$}$ & {$250$} & {$250$}& {$500$} & 500 &{$500$} & 500\\
\hline
$\textit{$M$}$ & {$2500$} & {$500$}& {$500$} & 333 & {$250$}& 333 \\
\hline
Code length & \multicolumn{6}{c|}{$10^5$}\\
\hline
{\tabincell{c}{Target Code rate}} & \multicolumn{3}{c|}{$0.5$}& 0.75&0.5&0.75 \\
\hline
$\text{Capacity limit}$ &  $0.110$ &  $0.572$  &  $1.206$  & 5.384 &   $2.669$ & 7.994\\
\hline
& \multicolumn{5}{c|}{\vspace{-0.35cm}}   &\\
\hline
{\tabincell{c}{Designed\vspace{-0.05cm}\\ Code rate}}  & $0.5000$  & $0.5013$  & $0.5029$ & $0.7370$ &$0.5021$& $0.7369$\\
\hline
${\textit{$R_{\cal C}$}}$ & $1.0000$  & $1.0026$  & $1.0058$ & $1.4741$ & $1.0042$& $1.4738$
\\
\hline
${\textit{R}}_{\text{sum}}$ & $249.99$  & $250.67$  & $502.90$ & 737.06 & $502.10$&736.91\\
\hline
 Iterations & 200  & 200 & 200 & 200 & 700 & 200\\
\hline
{\tabincell{c}{Check edge\vspace{-0.05cm}\\ distribution}} & ${\it{\eta}}_{\text{10}}=1$  & ${\it{\eta}}_{\text{9}}=1$  & ${\it{\eta}}_{\text{8}}=1$ & {\tabincell{c}{$\eta_{8}=0.5$ \vspace{-0.05cm}\\ $\eta_{20}=0.5$}} &${\it{\eta}}_{\text{7}}=1$& {\tabincell{c}{$\eta_{12}=0.8$ \vspace{-0.05cm}\\ $\eta_{80}=0.2$}} \\
\hline
  & $\lambda_2=0.1922$ & $\lambda_2=0.2254$ & $\lambda_2=0.2746$& $\lambda_2=0.5546$&$\lambda_2=0.4655$& $\lambda_2=0.4882$\\
& $\lambda_3=0.1694$ & $\lambda_3=0.2066$ & $\lambda_3=0.2622$ &$\lambda_{3}=0.1450$& $\lambda_3=0.1183$& $\lambda_{19}=0.3228$\\
 Variable&  $\lambda_7=0.2201$ & $\lambda_7=0.1101$ & $\lambda_{10}=0.2098$ & $\lambda_{40}=0.1750$ & $\lambda_{20}=0.1020$&$\lambda_{65}=0.0002$\\
 edge &  $\lambda_{8}=0.0511$   & $\lambda_{8}=0.1377$ & $\lambda_{40}=0.1950$ & $\lambda_{45}=0.1255$ &$\lambda_{21}=0.1827$&$\lambda_{67}=0.0002$\\
distribution   & $\lambda_{26}=0.0759$ & $\lambda_{27}=0.1294$ & $\lambda_{45}=0.0223$ &   &   $\lambda_{140}=0.1315$ &$\lambda_{100}=0.1201$\\
      & $\lambda_{27}=0.1315$ & $\lambda_{50}=0.0969$  & $\lambda_{90}=0.0361$ & &&$\lambda_{110}=0.0685$\\
       & $\lambda_{80}=0.0351$ & $\lambda_{60}=0.0939$  &  &   &&\\
      & $\lambda_{90}=0.1247$ &  &   & && \\
\hline
$snr^{\it{\ast}}_{\text{dB}}$ & $0.3$ & $0.69$ & $1.33$ & 5.62  & $2.77$ &8.5\\
\hline
\end{tabular}} 
\end{table*}
To verify the finite-length performance of the irregular LDPC codes with code rate $\approx0.5$, we provide the BER performances of the optimized codes. QPSK modulation is used. The rate of each symbol is $R_{\cal C}\approx1$ bits/symbol, and the sum rate is $R_{sum}\approx N$ bits per channel use. The maximum iteration number is $200\sim700$. Fig.~\ref{Fig:BER_AMP} shows that for all $\beta$, gaps between the BER curves of the codes at $10^{-5}$ and the corresponding Shannon limits are within $0.7 \sim 1$~dB.

To validate the advantage of matching principle, we provide AMP for a standard regular (3, 6) LDPC code (denoted as ``(3, 6)'') \cite{Gallager1962}, and a SISO irregular LDPC code \cite{Chung01} (denoted as ``SISO-Irreg''), corresponding to $R_{\mr{AMP-DEC}}$ discussed in Section \ref{Sec:Comparisons}. The parameters of ``SISO-Irreg'' are $\lambda(x)\!=\!0.170031x+0.160460x^2 +0.112837x^{5}  +0.047489x^6 +0.011481x^{9}\!+\!0.091537x^{10}  +0.152978x^{25} + 0.036131x^{26} +0.217056x^{99}$ and $\eta(x)\!=\!0.0625x^9 +0.9375x^{10}$, whose rate is $0.50004$ and decoding threshold is $0.0247$~dB away from the binary input AWGN capacity.

As shown in Fig.~\ref{Fig:BER_AMP}, when the BER curves of three systems are at $10^{-5}$, the optimized irregular LDPC codes  have $0.8 \sim 2$~dB performance gains over the un-optimized regular (3, 6) LDPC code for  $\beta=\{0.1, 0.5, 1, 2\}$, and $0.5 \sim 6$~dB performance gains over ``SISO-Irreg'' for $\beta=\{0.5, 1, 2\}$. For small $\beta$ (e.g. $\beta=0.1$), the ``SISO-Irreg'' is good enough, since the interference is negligible in this case (see Fig. \ref{Fig:area_expl}). These results demonstrate that code optimization provides attractive performance improvement, especially for the large $\beta$. \vspace{2mm}
 \begin{figure}[t]  
  \centering
  \includegraphics[width=7cm]{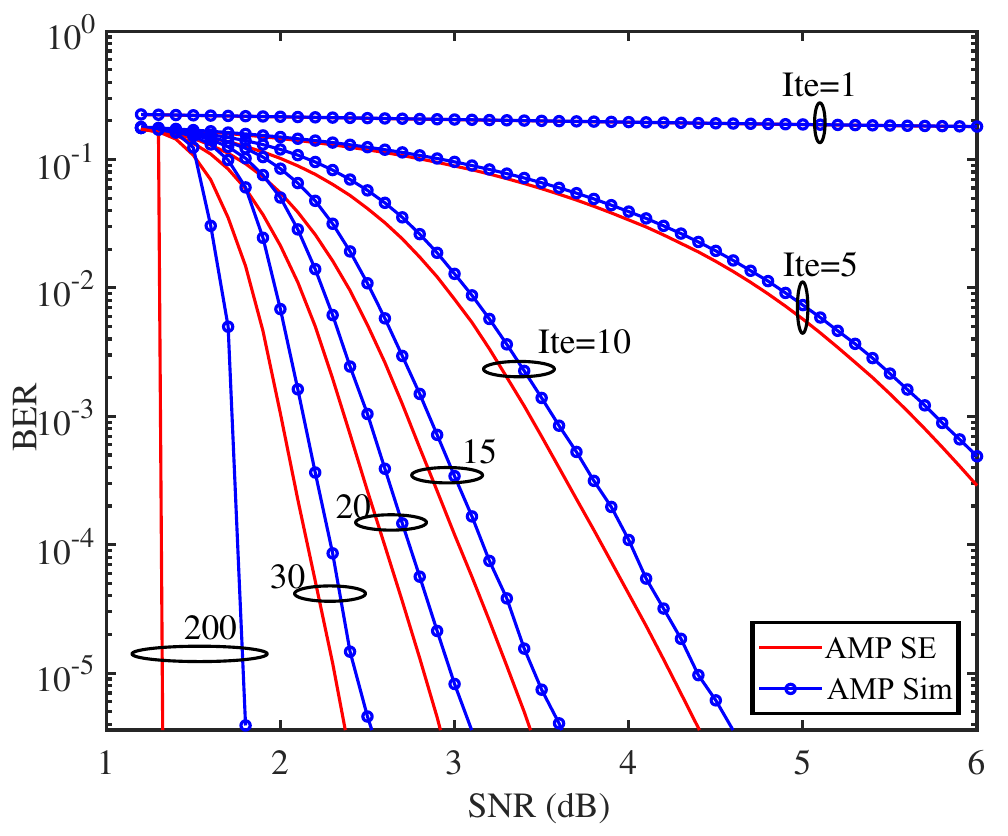}\\ 
  \caption{Comparison of simulation and SE predictions for AMP with optimized irregular LDPC code under QPSK modulation. The curves from right to left correspond to iterations $\mr{ite}=[1, 5, 10, 15, 20, 30, 200]$. Other parameters are the same as those of the case $\beta=1$ in Fig. \ref{Fig:BER_AMP} and Table \ref{Opt_degree1}.}\label{Fig:BER_AMP_conv} 
\end{figure} 
 
{\emph{Stability of Regular LDPC Code, SISO-Irregular LDPC}}
{\emph{Code and the Optimized Code:}} 
\begin{itemize}
    \item When $\beta$ is small, the curve of ``SISO-Irreg" matches the target curve $\omega^*_{\cal C}$ better than the regular LDPC code. Therefore, for $\beta=\{0.1,0.5\}$, ``SISO-Irreg" outperforms  the regular LDPC code. (See Fig. \ref{Fig:beta_match}(a).) 
     \item When $\beta$ is moderately large, the $(3, 6)$ regular LDPC code matches the target curve $\omega^*_{\cal C}$ better than ``SISO-Irreg". Therefore, for $\beta=\{1,2\}$, the regular LDPC code outperforms ``SISO-Irreg". (See Fig. \ref{Fig:beta_match}(b).) 
     \item The optimized code always has the best performance following the matching principle. 
\end{itemize} \vspace{2mm} 

{\emph{Number of Iterations:}} The number of iterations mainly depends on $\beta$. AMP has lower convergence speed when $\beta$ is large since the AMP LD  decreases more slowly. Consequently, it needs more iterations to converge. In our simulations, for $\beta=\{0.1, 0.5, 1,1.5\}$, 200 iterations is sufficient, while more iterations (e.g. 700) for higher $\beta$ (e.g. $\beta=2$). \vspace{2mm} 

{\emph{SE of AMP with LDPC Code:}} Fig. \ref{Fig:BER_AMP_conv} compares the simulated and predicted BER performances of AMP with optimized irregular LDPC code. As we can see, the SE predictions are accurate when the number of iterations is small (e.g. $\rm{Ite}\leq30$). The gap increases with the number of iterations. For $\mathrm{Ite}=200$, the simulated BER is about $0.5$ dB away from the SE curve.  \vspace{2mm}

\begin{figure*}[t]
  \centering
  \includegraphics[width=13cm]{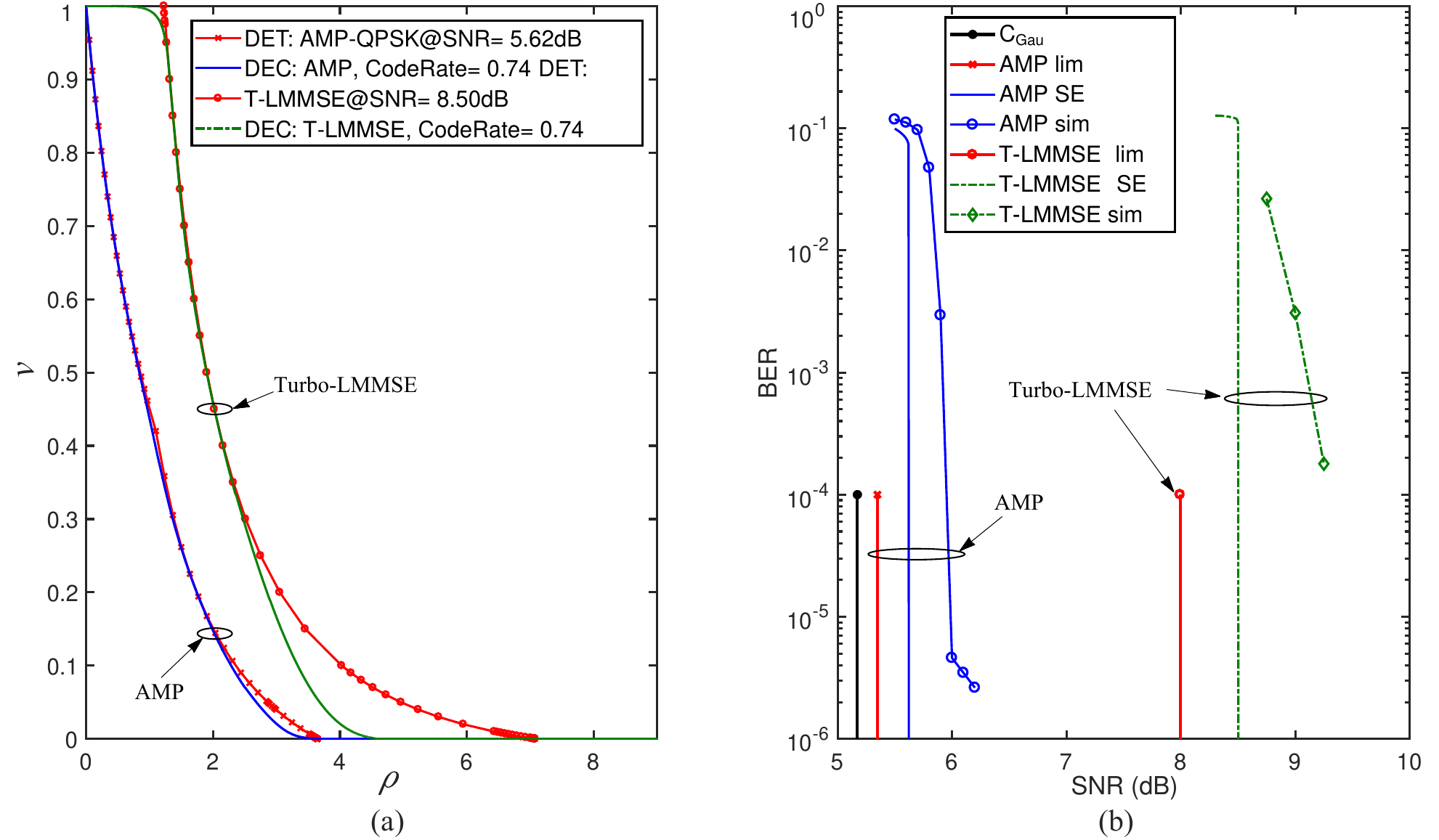}\\ 
  \caption{Transfer function matching (left) and BER performances (right) of AMP and Turbo-LMMSE  \cite{YC2018TWC, Lei20161b} with optimized irregular LDPC codes, where ``DET''  denotes the LD transfer function of AMP/Turbo-LMMSE, ``DEC''  the optimized decoding transfer function of AMP/Turbo-LMMSE, $C_{\rm Gau}$ the Gaussian capacity, ``SE''  the state evolution, ``lim''  the QPSK achievable rate limits of AMP/Turbo-LMMSE, ``sim''  the simulated BERs. Code length = $10^5$, code rate $\approx$ 0.74, QPSK modulation, iterations = $200$, and $\beta=1.5$ with $N=500$ and $M=333$, irregular LDPC codes are optimized for both AMP and Turbo-LMMSE. For more details, refer to Table \ref{Opt_degree1}.}\label{Fig:BER_AMP_Turbo} 
\end{figure*}

\subsection{BER Comparison with Optimized Turbo-LMMSE}
We now compare AMP and Turbo-LMMSE \cite{ YC2018TWC}. We consider a $500\times 333$ QPSK LRMS with $\beta=1.5$. As shown in Fig. \ref{Fig:Rate_AMP_Turbo}(b), the SNR limits of AMP and Turbo-LMMSE for the target rate $R_{\cal C}=1.48\approx1.5$ are $5.38$ dB and $7.99$ dB respectively. Fig. \ref{Fig:BER_AMP_Turbo}(a) shows the transfer functions of AMP and Turbo-LMMSE after optimization. The parameters of the optimized LDPC codes are listed in Table \ref{Opt_degree1}. The output of the DEC transfer function is an \emph{a-posteriori} variance, while that of Turbo is an \emph{extrinsic} variance. Fig. \ref{Fig:BER_AMP_Turbo}(b) shows the BER performances of AMP and Turbo-LMMSE (with iterations $=200$) using optimized LDPC codes. As we can see, the thresholds of AMP and Turbo-LMMSE are $5.62$ dB and $8.50$ dB respectively, 0.24 dB and 0.51 dB away from the corresponding achievable rate limits, and 0.6 dB and 1.2 dB away from their respective thresholds. We can see that, AMP has 3.5 dB improvement in BER  over Turbo-LMMSE.

{\emph{Complexity Comparison:}} The overall complexity of an iterative receiver including an LD and a DEC is $\mathcal{O}\left((\Xi_{LD}\!+\!\Xi_{DEC})N_{ite}\right)$, where $N_{ite}$ is the number of iterations, $\Xi_{LD}$ and $\Xi_{DEC}$ are complexities of LD and DEC per iteration respectively. For an LDPC decoder, $\Xi_{DEC}\approx 4\bar{d}_vN_c$, where $N_c$ is the code length and $\bar{d}_v=\big[\sum_{i}\lambda_i/i\big]^{-1}$ denotes the averaged variable-node degree. For AMP,  $\Xi_{LD}^{\mr{AMP}}=\mathcal{O}(MN)$. For Turbo-LMMSE,  $\Xi_{LD}^{\mr{Turbo}}=\mathcal{O}(MN^2)$ since it involves matrix inverse. Thus, AMP has much lower cost than Turbo-LMMSE.

\section{Conclusion}
This paper is on an AMP based scheme  for a coded LRMS with arbitrary input distributions. We show that AMP is information theoretically optimal using a curve matching principle and the IIDG assumption. In addition, a code design principle is provided for AMP, and the irregular LDPC codes are considered for binary signaling as an example. The numerical results show that AMP is capacity-approaching (i.e. within 1dB away from the limit) based on optimized irregular LDPC codes, and significant performance improvements ($0.8$ dB $\sim$ $4$ dB) are observed over the system without code optimization. Apart from that, AMP has lower complexity and better performance that the well-known Turbo-LMMSE algorithm.

The proof of Assumption \ref{Pro:SE_new} is an interesting future work. A rigorous SE proof for a certain kind of non-separable (e.g. uniformly Lipschitz) functions was established in \cite{Berthier2017}, which may be used to prove Assumption \ref{Pro:SE_new} in this paper.  
 
\appendices 

\section{Proof of Proposition \ref{The:area_LRMS}}\label{APP:Consistency} 
In this appendix, we will first show the consistency between the capacity given in \eqref{Eqn:dis_cap} and that derived in \cite{Reeves_TIT2019}. We then show $A_{\omega_{\cal C}^*}=C$, where $A_{\omega_{\cal C}^*}$ is given in \eqref{Eqn:dis_cap_new2} and $C$ in \eqref{Eqn:dis_cap}.

\subsection{Consistency Between   \texorpdfstring{\eqref{Eqn:dis_cap}}{TEXT} and the Results in \texorpdfstring{\cite{Reeves_TIT2019}}{TEXT}}  

Reference \cite{Reeves_TIT2019} is based on the following system model:
\BE\label{Eqn:linear_system_b}
\bf{y}=\bf{Ax}+\bf{n} 
\EE
where $\bf{A}$ is an IIDG matrix with $A_{ij}\sim \mathcal{N}({0},1/N)$,  $\bm{n}\!\sim\!\mathcal{N}(\mathbf{0}, \bm{I}_M)$, and $M,N\to\infty$ with a fixed $\delta=M/N$. Note that \eqref{Eqn:linear_system_b} is a real system, which is different from \eqref{Eqn:linear_system}. The capacity of \eqref{Eqn:linear_system_b} was derived in \cite{Reeves_TIT2019} as follows.

\begin{lemma}
Assume that $\zeta \!=\! \mr{mmse}_{X}\left(\delta/(1+\zeta)\right)$ has one positive fixed point $\zeta^*$. The capacity of the LRMS in \eqref{Eqn:linear_system_b} is given by
\BE\label{Eqn:C_Reeves}
C \!=\!  \frac{\delta}{2}\big[ \log({1+\zeta^*})-{\zeta^*}/({1+\zeta^*}) \big] + I\left(\delta/({1+\zeta^*})\right),
\EE 
where $I\left(s\right)\!=\!I(x;\sqrt{s}x+n)$,  $\mr{mmse}_{X}(s)=\mr{mmse}(x|\sqrt{s}x+n)$. 
\end{lemma} 

Note that the following:
\begin{enumerate}[(i)]
    \item In \eqref{Eqn:dis_cap}, ${1}/{2}$ is removed as \eqref{Eqn:linear_system} is a complex LRMS, while  \eqref{Eqn:linear_system_b} is a real one.
    \item Let $v_x$ be the variance of $x_i$ in \eqref{Eqn:C_Reeves}, we have 
    \BS\label{Eqn:eq_I_M}\begin{align} 
        &\mr{mmse}_{X}(s) = v_x\omega(v_x s), \\ 
        &I\left(s\right) =I\left(x; \sqrt{s} x+z)\right)  = C_{\rm SISO}(v_x s ).
    \end{align}
    \ES
    \item We have $ v_x= \delta^{-1}\sigma^{-2}=\beta snr$ with $\beta=\delta^{-1}=N/M$, since \eqref{Eqn:C_Reeves} considers $\bm{n}\!\sim\!\mathcal{N}(\mathbf{0}, \bm{I}_M)$ and $A_{ij}\sim \mathcal{CN}({0},1/N)$. Combining $ v_x=\beta snr$ and \eqref{Eqn:eq_I_M}, we have
\BS\begin{align}
    \mr{mmse}_{X}\big(\delta/(1\!+\!\zeta)\big) & =   \beta  snr  \,  \omega\big(snr/(1+\zeta)\big),\\
    I\left(\delta/({1+\zeta^*})\right) &= C_{\rm SISO}\big(snr/(1+\zeta^*)\big).
\end{align} \ES
\end{enumerate} 
Following (i)-(iii), we can see the consistency between the capacities in \eqref{Eqn:dis_cap} and \eqref{Eqn:C_Reeves}.

\subsection{Proof of Proposition \ref{The:area_LRMS}}\label{APP:Consistencyb} 
We now show $A_{\omega_{\cal C}^*}=C$, where $A_{\omega_{\cal C}^*}$ is given in \eqref{Eqn:dis_cap_new2} and $C$ in \eqref{Eqn:dis_cap}.

Let ${\rho}^*  \!=\! snr/(1\!+\!\zeta^*)$, i.e. $\zeta^*\!=\!snr/{\rho}^* \!-\!1$. Then the fixed point function in \eqref{Eqn:dis_cap} is rewritten to 
\BE
snr/{\rho}^* -1 = \beta  snr \,  \omega({\rho^*} ),
\EE
which is equivalent to the fixed point function $\omega(\rho)=\phi^{-1}(\rho)$. Substituting  \eqref{Eqn:C_mmse} and ${\rho}^*  = snr/(1+\zeta^*)$ into \eqref{Eqn:dis_cap_new2}, we have
\BS \begin{align}
 & A_{\omega_{\cal C}^*}  =  \beta^{-1}\big[\rho^{*}/snr\!-\!\log(\rho^{*}/snr)\!-\!1\big] \!+\!\! \int_0^{\rho^{*}} \!\!\! \omega(\rho) d \rho\\ 
&=  \beta^{-1}\left[ \log({1 \!+ \!\zeta^*})  - \!\frac{\zeta^*}{1\!+\!\zeta^*} \right] \!+  C_{\rm SISO} \left(\frac{snr}{ 1\!+\!\zeta^*}\right)\!.
\end{align}\ES 
This is the same as the capacity $C$ in \eqref{Eqn:dis_cap}. Hence, we complete the proof of Proposition \ref{The:area_LRMS}.

\section{An Alternative Proof of the Capacity of an LRMS}\label{APP:the_dis_cap} 
In this appendix, we provide an alternative proof for the capacity of an LRMS using the properties of AMP. We call $\mathcal{M}_x({snr}) \!\equiv\! \frac{1}{N}\mr{E}\big\{ \|\bf{x}\!-\!\hat{\bf{x}}_{\mr{MMSE}}\|^2 \big\}$ the MMSE of an LRMS and  $\mathcal{M}_{Ax}( {snr}) \equiv \tfrac{1}{N}\mr{E}\{ \|\bf{A}\bf{x}-\bf{A}\hat{\bf{x}}_{\mr{MMSE}}\|^2 \}$ the measurement MMSE of the LRMS. The following lemma gives the  capacity of an LRMS. 

\begin{lemma}[Measurement MMSE and Capacity]\label{The:dis_cap}
Assuming $\omega (\rho)= \phi^{-1} (\rho) $ has a unique positive solution $\rho^*$. The measurement MMSE of an LRMS is given by\vspace{-0.1cm}
\BS\BE\label{Eqn:relat_mmses_new}
\mathcal{M}_{Ax}({snr}) = \rho^*\omega(\rho^*)/snr= \rho^*\mathcal{M}_x({snr}) /snr,\vspace{-0.1cm}
\EE
and the capacity is given by
\BE\label{Eqn:dis_cap_new}
C  \!=\! A_{\omega_{\mathcal{C}}^*},
\EE\ES
where $A_{\omega_{\mathcal{C}}^*}$ is defined in \eqref{Eqn:dis_cap_new2}.
\end{lemma}

The next two subsections respectively give the proofs of \eqref{Eqn:relat_mmses_new} and \eqref{Eqn:dis_cap_new} in Lemma \ref{The:dis_cap}. 
\begin{itemize}
    \item In \ref{Sec:m_mmse}, we prove the measurement MMSE in \eqref{Eqn:relat_mmses_new} of an LRMS using the MMSE optimality \cite{Tulino2013, Barbier2017arxiv, Reeves_TIT2019} and the decoupling property  \cite{Takeuchi2019AMP} of AMP.
    \item In \ref{APP:proof_dis_cap}, we prove the capacity in \eqref{Eqn:dis_cap_new} of an LRMS based on the measurement MMSE in \eqref{Eqn:relat_mmses_new} using the vector-I-MMSE theorem \cite{Guo2005}. 
\end{itemize}

\subsection{Proof of the Measurement MMSE in \texorpdfstring{\eqref{Eqn:relat_mmses_new}}{TEXT}}\label{Sec:m_mmse}
As $t\to\infty$, the AMP in \eqref{Eqn:AMP} converges to  
\BE
\bf{s}^*= \bf{s}^\infty, \;\; \bf{r}^*= \bf{r}^\infty, \;\; \rho^*= \rho_\infty, \;\; \omega^* = \omega(\rho^*) =v_\infty.
\EE
If Assumption \ref{Pro:SCP} holds, from Lemma \ref{Lem:mmse}, we have $\bf{s}^*= \hat{\bf{x}}(\bf{y};{snr})$ and $\mathcal{M}_x({snr})  = \omega(\rho^*)$. The following proposition is proved for AMP in \cite{Takeuchi2019AMP} (see Theorem 1(A-a) in \cite{Takeuchi2019AMP}).
\begin{proposition}\label{Pro:div_indep}
Note that $\langle\eta'(\bf{r}^*)\rangle  = \rho^* \omega^*$. Define
\BS\label{Eqn:div_free}\begin{align}
\tilde{\bf{s}}  &=  \big[\omega^{*^{-1}}-\rho^* \big]^{-1}\big[{ \omega^{*^{-1}}\:{\bf{s}}^*  -  \rho^* \bf{r}^*}\big],\\
\tilde{\bf{z}}  &= \tilde{\bf{s}}- {\bf{x}},
\end{align}\ES
where the entries of $\tilde{\bf{z}}$ are IID with zero mean and variance $({\omega^{*^{-1}}-\rho^*})^{-1}$. Then, $\tilde{\bf{z}}$ can be treated as a random variable that is asymptotically independent\footnote{Let $\bf{A}=\bf{U\Lambda V}$. In \cite{Takeuchi2019AMP}, it is proved that the entries of $\bf{b}=\bf{V}\tilde{\bf{z}}$ are IIDG and independent with $\bf{n}$ and $\bf{U\Lambda}$. Based on this, substituting $\bf{A}=\bf{U\Lambda V}$ and $\bf{b}=\bf{V}\tilde{\bf{z}}$ into \eqref{Eqn:v_xa}, we obtain \eqref{Eqn:v_x}, which is the same as that $\tilde{\bf{z}}$ is independent with $\bf{n}$ and $\bf{A}$.} with $\bf{n}$ and $\bf{A}$.
\end{proposition}

Using Proposition \ref{Pro:div_indep}, we have the following lemma.

\begin{lemma} \label{Pro:MMSE_of_A}
The following equation asymptotically holds for the fixed point of AMP:
 \begin{align}\label{Eqn:MMSE_of_A}
 \tfrac{1}{N}\mr{E}\big\{\!(\bf{x}\!-\!{\bf{s}^*})(\bf{x}\!-\!{\bf{s}^*})^{\rm H}\big\} \!=\! \tfrac{1}{N} \!\big[\big(\omega^{*^{-1}}\!\!-\!\rho^*\big)\bf{I} \!+\!  {snr}\bf{A}^{\rm H}\!\bf{A} \big]^{-1}\!.
\end{align}
\end{lemma}

\begin{IEEEproof}  
We rewrite \eqref{Eqn:div_free} in Proposition \ref{Pro:div_indep} as
\BE\label{Eqn:NLD_r}
\bf{r}^*  = \frac{\bf{s}^*}{\rho^*\omega^*} + \Big(1- \frac{1}{\rho^*\omega^*}\Big)\tilde{\bf{s}}.
\EE
The output of AMP LD in \eqref{Eqn:AMP} converges to
\BE\label{Eqn:AMP_fixed_point}
\bf{r}^*=\bf{s}^* + \bf{A}^{\rm H}(\bf{y}-\bf{A}\bf{s}^*) + \beta \langle\eta'(\bf{r}^{*})\rangle (\bf{r}^{*}-\bf{s}^{*}).
 \EE
For MMSE function $\eta$ \cite{Bayati2011},
\BE\label{Eqn:mmse_deriv}
\langle\eta'(\bf{r}^*)\rangle    = \rho ^* \omega^* .
\EE
With \eqref{Eqn:mmse_deriv} and the fixed point function ${snr} = {\rho^*}\big[{1-\beta\rho^*\omega(\rho^*)}\big]^{-1}$, \eqref{Eqn:AMP_fixed_point} can be rewritten to
\begin{align}\label{Eqn:AMP_fix_eqs}
&\bf{r}^*=\big[\bf{I} -  {snr}\bf{A}^{\rm H}\bf{A}/{\rho^*} \big] \bf{s}^* +    {snr}\bf{A}^{\rm H}\bf{y}/{\rho^*}.
\end{align}
From \eqref{Eqn:NLD_r} and \eqref{Eqn:AMP_fix_eqs}, we have
\BE
\bf{B} {\bf{s}^*} =  {snr}\bf{A}^{\rm H}\bf{y} + \big(\omega^{*^{-1}}-\rho^*\big)\tilde{\bf{s}},
\EE
where $\bf{B}=\big(\omega^{*^{-1}}-\rho^*\big)\bf{I} +  {snr}\bf{A}^{\rm H}\bf{A} $.
Thus,
\BE\label{Eqn: mmse_x}
 {\bf{s}^*} =  \bf{B}^{-1}  \big[{snr}\bf{A}^{\rm H}\bf{y} + \big(\omega^{*^{-1}}-\rho^*\big)\tilde{\bf{s}}\big].
\EE
Substituting  \eqref{Eqn:div_free} and $\bf{y}=\bf{Ax}+\bf{n}$ into \eqref{Eqn: mmse_x}, we have
\BS\begin{align}
&\tfrac{1}{N}\mr{E}\big\{ (\bf{x}-{\bf{s}^*})(\bf{x}-{\bf{s}^*})^{\rm H} \big\}\\
&= \tfrac{1}{N} \Big[\bf{B}^{-1}[{snr}\bf{A}^{\rm H}\bf{n} + \big(\omega^{*^{-1}}-\rho^*\big)\tilde{\bf{z}}]\Big]^2  \label{Eqn:v_xa}\\
&= \tfrac{1}{N}\big[\big(\omega^{*^{-1}}-\rho^*\big)\bf{I} +  {snr}\bf{A}^{\rm H}\bf{A} \big]^{-1}\label{Eqn:v_x},
\end{align}\ES
where \eqref{Eqn:v_x} follows Proposition \ref{Pro:div_indep}.
Thus, we complete the proof of Lemma~\ref{Pro:MMSE_of_A}.
\end{IEEEproof}

Based on Lemma \ref{Pro:MMSE_of_A}, the MSE of AMP is given by
 \BS\label{Eqn:SE_FP} \begin{align}
\mathcal{M}_x(snr)& =\tfrac{1}{N}\mr{E}\big\{\|\bf{x}-{\bf{s}^*}\|^2\big\}=\omega(\rho^*)\\
&= \tfrac{1}{N}\mr{Tr}\Big\{ \big[\big(\omega^{*^{-1}}-\rho^*\big)\bf{I} +  {snr}\bf{A}^{\rm H}\bf{A} \big]^{-1}\Big\}\\
&= \mr{E}_{\lambda_{\bf{A}^{\rm H}\bf{A}}}\Big\{\!\big[\big(\omega^{*^{-1}}\!\!-\!\rho^*\big) +  {snr}\lambda_{\bf{A}^{\rm H}\bf{A}}\big]^{-1} \!\Big\},
\end{align}\ES
where ${\lambda_{\bf{A}^{\rm H}\bf{A}}}$ is the eigenvalue of ${{\bf{A}^{\rm H}\bf{A}}}$. Also, the measurement MMSE is derived as
\BS\begin{align}
&\mathcal{M}_{Ax}({snr})\\ 
& =\tfrac{1}{N}\mr{E}\big\{\|\bf{A}\bf{x}-\bf{A}{\bf{s}^*}\|^2\big\} \\
& =\tfrac{1}{N}\mr{Tr}\big\{\|\bf{A}(\bf{x}-{\bf{s}^*})(\bf{x}-{\bf{s}^*})^{\rm H}\bf{A}^{H} \big\} \\
& = \tfrac{1}{N} \mr{Tr}\Big\{\bf{A}\big[(\omega^{*^{-1}}-\rho^*)\bf{I} +  {snr}\bf{A}^{\rm H}\bf{A} \big]^{-1}\bf{A}^{H} \Big\}\label{Eqn:v_ax1}\\
& = \mr{E}_{\lambda_{\bf{A}^{\rm H}\bf{A}}}\Big\{ \lambda_{\bf{A}^{\rm H}\bf{A}}\big[(\omega^{*^{-1}}-\rho^*) +  {snr}\lambda_{\bf{A}^{\rm H}\bf{A}}\big]^{-1} \Big\}\\
& =\! snr^{\!-1}\!\left[1\!-\! \mr{E}_{\lambda_{\bf{A}^{\rm H}\bf{A}}} \Big\{\!\big[1 \!+\!  {snr}(\omega^{*^{-1}}\!\!\!\!-\!\rho^*)^{-1}\lambda_{\bf{A}^{\rm H}\bf{A}}\big]^{\!-1} \!\Big\}\!\right]\\
&= \rho^*\omega(\rho^*)/snr \label{Eqn:v_ax_a}\\
&= \rho^*\mathcal{M}_x({snr}) /snr, \label{Eqn:v_ax_b}
\end{align}\ES
where \eqref{Eqn:v_ax1} follows \eqref{Eqn:MMSE_of_A}, and \eqref{Eqn:v_ax_a} and \eqref{Eqn:v_ax_b} follows \eqref{Eqn:SE_FP}. Therefore, we obtain  \eqref{Eqn:relat_mmses_new}.

\subsection{Proof of the Capacity in \texorpdfstring{\eqref{Eqn:dis_cap_new}}{TEXT}}\label{APP:proof_dis_cap}
The connection between the measurement MMSE and the capacity of an LRMS is given by Lemma \ref{Lem:V-I-MMSE} proven in \cite{Guo2005}.

\begin{lemma}[Vector I-MMSE]\label{Lem:V-I-MMSE}
Consider a system $\bf{y}=\sqrt{snr}\bf{A}\bf{x}+\bf{z}$ where $\bf{x}\sim P_{\bf{x}}$ and $\bf{z}\sim \mathcal{CN}(\bf{0},\bf{I})$. Then, the capacity of this system is given by
\BE\label{Eqn:dis_C}
C = \tfrac{1}{N}I(\bf{x}; \sqrt{{snr}}\bf{A}\bf{x}+\bf{z}) =\int_0^{snr} \!\!\!\mathcal{M}_{Ax}(\rho)\:d\:\rho.
\EE 
\end{lemma}

From \eqref{Eqn:relat_mmses_new} and Lemma \ref{Lem:V-I-MMSE}, we have
\BS\begin{align}
C &=  \int_0^{snr}\mathcal{M}_{Ax}({s}) \:d\: {s}\label{Eqn:disC_a}\\
&=  \int_0^{snr} \rho\omega(\rho)/s \:d\: {s}\label{Eqn:disC_b} \\
&=  \int_0^{\rho^*} [{1-\beta\rho\omega(\rho)}] \omega(\rho) \:d\: \frac{\rho}{1-\beta\rho\omega(\rho)},\label{Eqn:disC_c}
\end{align}\ES
where \eqref{Eqn:disC_a} follows \eqref{Eqn:dis_C}, \eqref{Eqn:disC_b} follows \eqref{Eqn:relat_mmses_new}, and \eqref{Eqn:disC_c} follows the fixed point function. The following manipulations show that $C=A_{\omega_{\mathcal{C}}^*}$.
\BS\begin{align}
C&= \int_0^{\rho^*}  \frac{\omega+\beta\rho^{2}\omega\omega'}{1-\beta\rho\omega } \:d\: \rho\\
&= \int_0^{\rho^*}  \frac{\omega+\rho\omega'-\rho\omega'(1-\beta\rho\omega)}{1-\beta\rho\omega } \:d\: \rho \\
&= -\int_0^{\rho^*} \rho  \:d\: \omega +
 \int_0^{\rho^*}  \frac{\omega+\rho\omega'}{1-\beta\rho\omega } \:d\: \rho \\
&= \int_0^{\rho^*} \rho^{*}  \:d\: \omega + \beta^{-1}\int_0^{\rho^*}  \:d\: \log(1-\beta\rho\omega) \\
&= \left[-{\rho\omega} - \beta^{-1}\log(1-\beta\rho\omega) \right]_{\rho=0}^{\rho=\rho^*} + \!\int_0^{\rho^*} \!\!\! \omega d \rho \label{Eqn:disC_i}\\ 
&=\beta^{-1}[\rho^{*}/snr-\log(\rho^{*}/snr)-1] + \!\int_0^{\rho^{*}} \!\!\!\omega(\rho) d \rho \\
&=A_{\omega_{\mathcal{C}}^*},\label{Eqn:disC_j}
\end{align}\ES
where $\omega'$ is the derivative of $\omega(\rho)$, \eqref{Eqn:disC_i} follows the fixed point function, and \eqref{Eqn:disC_j} follows \eqref{Eqn:dis_cap_new2}. Thus, we obtain \eqref{Eqn:dis_cap_new}.

\section{Gaussian Signaling}   
We now study a special case of Gaussian $\bf{x}$. In \ref{APP:Gau_Instance}, we will show a simpler proof as well as a closed-form expression of the area property. In \ref{APP:Gau_SCM}, we will show the curve matching condition asymptotically holds, i.e., there exists an $n$-layer SCM code whose transfer function matches with the desired $\omega_{{\mathcal{C}-{\mr{Gau}}}}^*(\rho)$ as $n\to \infty$. 

\subsection{Gaussian Area Property} \label{APP:Gau_Instance}
The following lemma gives an explicit form of the measurement MMSE for Gaussian signaling.

\begin{lemma}[Gaussian Measurement MMSE]\label{lem:gau_mmse}
For Gaussian signaling, the measurement MMSE of an LRMS is given by\vspace{-0.1cm}
\BE\label{Eqn:relat_mmses_Gau}
\mathcal{M}_{Ax}({snr}) = snr^{-1} \rho^*_{\mr{Gau}}/(1+\rho^*_{\mr{Gau}}),\vspace{-0.1cm}
\EE
where $\rho^*_{\mr{Gau}}$ is given in \eqref{Eqn:Gau_rho}.
\end{lemma}
\begin{IEEEproof} 
For $\bf{x}\sim\mathcal{CN}(\bf{0},\bf{I})$, the following LMMSE detection is a global MMSE estimation of an uncoded LRMS.
\BE
\bf{s}^* = (\bf{A}^{\rm H}\bf{A}^{\rm H}+snr^{-1}\bf{I})^{-1}\bf{A}^{\rm H}\bf{y}.
\EE
Its average MSE is the corresponding MMSE, i.e.,
\BS\begin{align}
\mathcal{M}_x(snr) &= \tfrac{1}{N}\mr{Tr}\big\{(\bf{s}^*-\bf{x})(\bf{s}^*-\bf{x})^{\rm H}\big\}\\ &=\tfrac{1}{N}\mr{Tr}\big\{(snr\bf{A}^{\rm H}\bf{A}+\bf{I})^{-1} \big\}\\
 &= \mr{E}_{\lambda_{\bf{A}^{\rm H}\!\bf{A}}}\Big\{\big(1 +  {snr}\lambda_{\bf{A}^{\rm H}\bf{A}}\big)^{-1} \Big\}.   
\end{align} 
\ES
Then, we obtain the measurement MMSE as
\BS\begin{align}
\mathcal{M}_{Ax}(snr) &= \tfrac{1}{N}\mr{Tr}\big\{\bf{A}(\bf{s}^*-\bf{x})(\bf{s}^*-\bf{x})^{\rm H}\bf{A}^{\rm H}\big\}\\
&= \tfrac{1}{N}\mr{Tr}\big\{\bf{A}^{\rm H}\bf{A}(snr\bf{A}^{\rm H}\bf{A}^{\rm H}+\bf{I})^{-1}\big\}\\
&= \mr{E}_{\lambda_{\bf{A}^{\rm H}\bf{A}}}\Big\{\lambda_{\bf{A}^{\rm H}\bf{A}}\big(1 +  {snr}\lambda_{\bf{A}^{\rm H}\bf{A}}\big)^{-1} \Big\}\\
& = snr^{-1}\big(1-\mathcal{M}_x(snr)\big).
\end{align}\ES
In addition, according to \eqref{Eqn:snr_rho} and $\mathcal{M}_x(snr) =\omega_{\mr{Gau}}(\rho^*_{\mr{Gau}})=1/(1+\rho^*_{\mr{Gau}})$, we have
\BS\label{Eqn:Gau_mMMSE} \begin{align} 
\mathcal{M}_{Ax}(snr) &=\rho^*_{\mr{Gau}} \mathcal{M}_x(snr)/snr \\
&=snr^{-1} \rho^*_{\mr{Gau}}/(1\!+\!\rho^*_{\mr{Gau}}).
\end{align}
 \ES
Thus, we obtain Lemma \ref{lem:gau_mmse}. 
\end{IEEEproof}

Now we are ready to prove \eqref{Eqn:dis_cap_Gau}. The Gaussian capacity of an LRMS is  \cite{David2005} 
\BS\label{Eqn:capacity}\begin{align}
&C_{\mr{Gau}}=\tfrac{1}{N}I(\bf{x};\bf{y})
= \tfrac{1}{N}\log \det(\bf{I} +  snr\bf{A}^{\rm H}\bf{A}),
\end{align}
which is achieved by $\bf{x}\sim\mathcal{CN}(\bf{0},\bf{I})$.  For IIDG $\bf{A}$ with $A_{ij}\sim \mathcal{CN}({0},1/M)$, we have \cite{Tulino2004}
 \begin{align}
C_{\mr{Gau}} \to& \log[1+snr- \mathcal{F}] \nonumber\\
& \;\; + \beta^{-1}\log[1+snr\beta- \mathcal{F}]-snr^{-1}\beta^{-1}  \mathcal{F},
\end{align}\ES
where $\mathcal{F}\!=\!0.25\left(\!\!\sqrt{snr(1\!+\!\sqrt{\beta })^2\!+\!1} -\!\sqrt{ snr(1\!-\!\sqrt{\beta})^2\!+\!1}\right)^2\!$.

Next, we show that $A_{\omega_{\mathcal{C}-{\mr{Gau}}}^*}$ is equal to the Gaussian capacity $C_{\mr{Gau}}$. From \eqref{Eqn:Gau_area},  
\BS\begin{align}
A_{\omega_{\mathcal{C}-{\mr{Gau}}}^*}  
\!\!=\!\beta^{-1}\log(1\!+\!\beta snr\,v^{*}_{\mr{Gau}})\!-\!\log(v^{*}_{\mr{Gau}})\!+\!v^{*}_{\mr{Gau}}\!\!-\!1,
\end{align}\ES
with
\BS\begin{align}
 v^{*}_{\mr{Gau}} &\!=  \phi^{-1}(\rho^{*}_{\mr{Gau}})\\
& \!=  \frac{\beta\!-\!1\!-\!snr^{-1}\!+\!\sqrt{\!(\beta\!-\!1)^2\!+\!2(\beta\!+\!1)snr^{-1}\!+\!snr^{-2}}}{2\beta}\\
& \!= 1\!-\!snr^{-1}\mathcal{F}/\beta.
\end{align}\ES
Therefore,\vspace{-0.3cm}
\BS\label{Eqn:rat_amp}\begin{align}
 A_{\omega_{\mathcal{C}-{\mr{Gau}}}^*}   
&\!\!=\!-{\log\left(\!1\!-\!\frac{\mathcal{F}}{\beta snr}\!\right)} \!+\! \frac{1}{\beta}\log(1\!+\!\beta snr\!-\!\mathcal{F})\!-\!\frac{snr\mathcal{F}}{\beta}\\
&\!=\!\log(1\!+\!snr\!-\!\mathcal{F}) \!+\!\frac{1}{\beta}\log(1\!+\!\beta snr\!-\!\mathcal{F})\!-\!\frac{\mathcal{F}}{\beta snr},
\end{align}\ES
where the second equation follows from $(1-snr^{-1}\mathcal{F}/\beta)(1+snr-\mathcal{F})=1$. From \eqref{Eqn:rat_amp} and \eqref{Eqn:capacity}, we have \eqref{Eqn:dis_cap_Gau}:
\BE
C_{\mr{Gau}} = A_{\omega_{\mathcal{C}-{\mr{Gau}}}^*}.
\EE

\subsection{Code Existence for Gaussian Signaling}\label{APP:Gau_SCM}
It was proved in \cite{Yuan2014} that there exists an infinite-layer SCM code whose extrinsic decoding transfer function asymptotically approaches a target monotonically decreasing curve. In this subsection, we extend this result to the APP decoding transfer function. We will  show that there is an SCM code whose transfer function $\omega_{C_n}(\rho)$ asymptotically approaches $\omega_{\mathcal{C}-\mr{Gau}}^*$. Then, we will then show that  the rate of this SCM code asymptotically approaches to $A_{\omega_{\mathcal{C}-\mr{Gau}}^*}$.

We first show that $R_{\mr{AMP}}$ can be approached with an infinite-layer SCM code. It is easy to verify that ${\phi}^{-1}({\rho})$ satisfies the following regularity conditions:
\begin{enumerate}[(a)]
\item   ${\phi^{-1}}(\rho) \geq0$, for $\rho\in[0,snr]$;
\item  monotonically decreasing in $\rho\in[0,\infty)$;
\item  equation $(p^{-1}+\rho) {\phi^{-1}}(\rho)=1$ has only one positive solution $\rho^*_p$ for any $p\in(0, 1]$;

\item continuous and differentiable in $[\rho^*_{\mr{Gau}},\infty)$ except for a countable set of values of $\rho$. 
\end{enumerate}

Consider an $n$-layer SCM code $x=\sum_{i=0}^{n-1}x_i$ and the power of $x_i$ is $p_{x_i}={1}/{N}$. In addition, $x_i$ is encoded using an idea random code with code rate\footnote{Note that \eqref{Eqn:R_i} considers Gaussian signaling for each $x_i$. For discrete signaling, from Lemma 1 in \cite{Guo2005}, we can complete the proof by replacing \eqref{Eqn:R_i} with
\BE\label{Eqn:R_i_discrete}
R_{n,i}= \frac{1/n}{{\rho^{*^{-1}}_{1-i/n}}+(n-i-1)/n} + o\left(\frac{1/n}{{\rho^{*^{-1}}_{1-i/n}}+(n-i-1)/n} \right),\nonumber
\EE
for all $i\in\{0,\cdots,n-1\}$.}
\BE\label{Eqn:R_i}
R_{n,i}= \log\left(1+ \tfrac{1/n}{{\rho^{*^{-1}}_{1-i/n}}+(n-i-1)/n}\right),
\EE
where $i\in\{0,\cdots,n-1\}$, and $\rho^*_{i/n}$ is the positive solution of $(\rho+n/i)\phi^{-1}(\rho)=1$. Condition (c) ensures the existence of $\{\rho^*_{i/n}\}$, and condition (b) ensures $\rho^*_{1}<\dots<\rho^*_{2/n}<\rho^*_{1/n}$.

For any $i\in\{0,\dots, n-1\}$ and the decoder's input $\rho\in[\rho^*_{1-i/n},\rho^*_{1-(i+1)/n})$, the first $i+1$ layers $[x_0,\cdots,x_{i}]$ can be successively decoded in the order from $x_0$ to $x_i$. Thus, under APP decoding, the transfer function of $x=\sum\limits_{i=0}^{n-1}x_i$ for input $x+\rho^{-1/2}z$ with $z \sim\mathcal{CN}(0,1)$ is  
\BE
\omega_{C_n}(\rho)\!=  \! \left\{ \!\!\!\begin{array}{l}
\frac{1}{\rho+1}, \qquad \quad 0 \le {\rho } < \rho^*_{1}\\
\frac{1}{\rho+n/(n-i)}, \quad\! \rho^*_{1-i/n} \le {\rho } < \rho^*_{1-(i+1)/n} \\
0, \qquad \qquad\;\;\rho^*_{1/n}< {\rho }< \infty
\end{array} \right., 
\EE
where $i= 1,\cdots,n-2$.

\begin{figure}[t]  
  \centering
  \includegraphics[width=7cm]{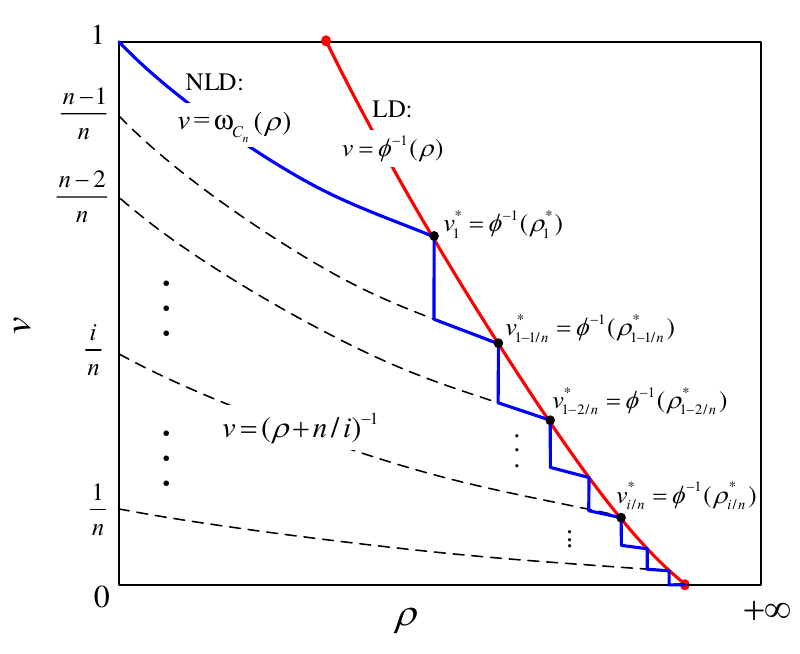}\\  
  \caption{An illustration of the transfer function of AMP LD $v=\phi^{-1}(\rho)$ and the corresponding transfer function $v=\omega_{C_n}(\rho)$ of the asymptotically matched $n$-layer SCM decoder.}\label{Fig:matched_curves} 
\end{figure}
Fig. \ref{Fig:matched_curves} shows the transfer functions of $n$-layer-SCM decoder ($v=\omega_{\mathcal{C}_n}(\rho)$) and LD ($v={\phi^{-1}}(\rho)$). Conditions (a)-(c) ensure that the decoder transfer function lies below that of LD, i.e.,
\BE
\omega_{\mathcal{C}_n}(\rho)\leq{\phi^{-1}}(\rho), \;\;\;\forall \rho\geq0.
\EE

Define $v\equiv f(\rho)=[1/\phi^{-1}(\rho)-\rho]^{-1}$, and we have $\rho^*_v=f^{-1}(v)$, where $f^{-1}(\cdot)$ is the inverse function of $f(\cdot)$. Then, as $n\to\infty$, the sum rate of the SCM code is given by
\BS\begin{align}
\! R_n 
 \!&= \lim\limits_{n\to\infty} \;\sum\limits_{i=0}^{n-1} \log\left(1+ \frac{1/n}{{\rho^{*^{-1}}_{1-i/n}}+(n-i-1)/n}\right)\\
& = \lim\limits_{n\to\infty} \; \frac{1/n}{{\rho^{*^{-1}}_{1-i/n}}+(n-i-1)/n}\\
& = \int_{0}^1 [{\rho^{*^{-1}}_{v}+v}]^{-1}d v \\
& = \int_{0}^1 \big[{[f^{-1}(v)]^{-1}+v}\big]^{-1}d v\\
& \!= \!\left[\!\big[ \rho^{-1}\!\!+\!f(\rho)\big]^{\!-1}\!f(\rho)\!\right]_{\rho=\rho^*_0}^{\rho=\rho^*_1} \!+\!\! \int_{\rho^*_1}^{\rho^*_0}\!\!\!f(\rho)  d  {[\rho^{-1}\!+\!f(\rho)]^{-1}} \label{Eqn:pro1}\\
& \!= \!\!\left[\rho\phi^{-1}(\rho)\right]_{\rho=\rho^*_0}^{\rho=\rho^*_1} \!+\!\! \int_{\rho^*_1}^{\rho^*_0}\!\!\frac{\phi^{-1}(\rho)}{1\!-\!\rho\phi^{-1}(\rho) }\: d\, {\rho\big(1\!-\!\rho\phi^{-1}(\rho)\big)} \label{Eqn:pro2}\\ 
& =  \left[\log(1-\rho\phi^{-1}(\rho))\right]_{\rho=\rho^*_1}^{\rho=\rho^*_0}  + \int_{\rho^*_1}^{\rho^*_0}  \phi^{-1}(\rho) \: d \rho \\
& =  \log(1+\rho^*_1)+ \int_{\rho^*_1}^{\infty}  \phi^{-1}(\rho) \: d \rho\label{Eqn:pro4} \\
& = A_{\omega_{\mathcal{C}-\mr{Gau}}^*},\label{Eqn:pro5}
\end{align}\ES
where \eqref{Eqn:pro1} follows the \emph{inverse integral lemma} below
\BS\begin{align}
&\int g\big(y,f^{-1}(y)\big) dy\\
&= g\big(f(x),x\big)f(x)-\!\! \int \!\!\! f(x)\,{d}\, g\big(f(x),x\big) + {\rm Constant},
\end{align}\ES
 \eqref{Eqn:pro2} from $ f(\rho)=[1/\phi^{-1}(\rho)-\rho]^{-1}$, \eqref{Eqn:pro5} follows \eqref{Eqn:dis_cap_new2} and $\omega_{\mr{Gau}}(\rho)=1/(1+\rho)$, and \eqref{Eqn:pro4} utilizes the following facts:
\begin{itemize}
\item  $ 1-\rho^*_1\phi^{-1}(\rho^*_1)=1/(1+\rho^*_1)$;
\item $\rho^*_0\phi^{-1}(\rho^*_0)=0$ from $\phi^{-1}(\rho^*_0)=0$ if $\rho^*_0$  is finite;
\item $\phi^{-1}(\rho)=0$ for any $\rho> \rho^*_0$ if $ \rho^*_0$ is finite, since $\phi^{-1}(\rho_1^*)=0$, and $\phi^{-1}(\rho)$ is positive and monotonically decreasing in $\rho\in[0,\infty)$.
\end{itemize}
Thus, we obtain the desired result:  $R_n=A_{\omega_{\mathcal{C}-\mr{Gau}}^*}$.

\end{document}